\documentclass[12pt,preprint]{amsart}

\usepackage{amsmath, latexsym, amsfonts, amssymb, amsthm}
\usepackage{graphicx, color,hyperref,dsfont,epsfig,caption,wrapfig,subfig}
\usepackage{pstricks,yfonts,mathalfa}
\usepackage{mathtools}
\usepackage{movie15}
\usepackage{soul}
\setstcolor{red}

\hypersetup{
    linktoc=page,
    linkcolor=red,          % color of internal links
    citecolor=blue,        % color of links to bibliography
    filecolor=blue,      % color of file links
    urlcolor=cyan,
    colorlinks=true           % color of external links
}

\numberwithin{equation}{section}

\newtheorem{theorem}{Theorem}[section]
\newtheorem{lemma}[theorem]{Lemma}

\newtheorem{corollary}[theorem]{Corollary}

\newtheorem{remark}[theorem]{Remark}
\newtheorem{definition}[theorem]{Definition}

\theoremstyle{definition}

\renewcommand{\tilde}{\widetilde}          % wider `tilde'
\DeclareMathSymbol{\leqslant}{\mathalpha}{AMSa}{"36} % nicer `smaller or equal'
\DeclareMathSymbol{\geqslant}{\mathalpha}{AMSa}{"3E} % nicer `larger or equal'
\DeclareMathSymbol{\eset}{\mathalpha}{AMSb}{"3F}     % nicer `emptyset'
\renewcommand{\leq}{\;\leqslant\;}                   % redef. of < or =
\renewcommand{\geq}{\;\geqslant\;}                   % redef. of > or =
\newcommand{\dd}{\text{\rm d}}             % a straight d for differentials

\newcommand{\C}{\mathbb{C}}
\newcommand{\R}{\mathbb{R}}
\newcommand{\Z}{\mathbb{Z}}

\newcommand{\eqlaw}{\overset{\text{(law)}}{=}}

\newcommand{\E}{\mathds{E}}

\newcommand{\ps}[1]{\langle #1 \rangle}

\def\eps{\varepsilon}
\def\S{\mathbb{S}}

\def\bi{\begin{itemize}}
\def\ei{\end{itemize}}

\def\bnum{\begin{enumerate}}
\def\enum{\end{enumerate}}
\def\<#1{\langle #1 \rangle}

\definecolor{ao(english)}{rgb}{0.0, 0.5, 0.0}
\definecolor{darkcandyapplered}{rgb}{0.64, 0.0, 0.0}
\definecolor{amber}{rgb}{1.0, 0.49, 0.0} 	
\definecolor{auburn}{rgb}{0.2, 0.2, 0.6}

\def\green#1{#1}

\newcommand{\norm}[1]{\left\lvert#1\right\rvert}
\newcommand{\expect}[1]{\mathbb{E}\left[#1\right]}
\usepackage{bm}
\usepackage{bbm}

\title[Toda Conformal Field Theories]{Probabilistic construction of Toda Conformal Field Theories}

\author{Baptiste Cercl\'e}
\email{baptiste.cercle@universite-paris-saclay.fr}
\address{Laboratoire de Math\'ematiques d'Orsay, B\^atiment 307. Facult\'e des Sciences d'Orsay, Universit\'e Paris-Saclay. F-91405 Orsay Cedex, France}
\author{R\'emi Rhodes }
\email{remi.rhodes@univ-amu.fr}
\address{Aix-Marseille University, Institut de Math\'ematiques de Marseille (I2M), and Institut Universitaire de France (IUF)}
\author{Vincent Vargas}
\email{vargas@math.univ-paris13.fr}
\address{LAGA Universit\'e Sorbonne Paris Nord, CNRS, UMR 7539, F-93430, Villetaneuse, France}

\begin{document}

\maketitle
\begin{abstract}
    Following the 1984 seminal work of Belavin, Polyakov and Zamolodchikov on two-dimensional conformal field theories, Toda conformal field theories were introduced in the physics literature as a family of two-dimensional conformal field theories that enjoy, in addition to conformal symmetry, an extended level of symmetry usually referred to as \textit{W-symmetry} or \textit{higher-spin} symmetry. More precisely Toda conformal field theories provide a natural way to associate to a finite-dimensional simple and complex Lie algebra a conformal field theory for which the algebra of symmetry \textit{contains} the Virasoro algebra. In this document we use the path integral formulation of these models to provide a rigorous mathematical construction of Toda conformal field theories based on probability theory. By doing so we recover expected properties of the theory such as the Weyl anomaly formula with respect to the change of background metric by a conformal factor and the existence of Seiberg bounds for the correlation functions.
\end{abstract}
 
\section{Introduction}
%%%%%%%%%%%%%%%%%% 
 \subsection{Toda Conformal Field Theories in the physics literature}
 In 1981, Polyakov presented  in a pioneering work \cite{Pol81} a canonical way of defining the notion of random surface, usually called Liouville conformal field theory (Liouville CFT hereafter), which is now considered to be an essential feature in the understanding of non-critical string theory and two-dimensional quantum gravity. A few years later Belavin, Polyakov and Zamolodchikov (BPZ) presented in their 1984 seminal work \cite{BPZ} a systematic procedure to solve models which like Liouville CFT possess certain conformal symmetries,  now referred to as \textit{two-dimensional conformal field theories} (CFTs in the sequel). The main input of their method was to exploit the constraints imposed by conformal symmetry through the study of the algebra of its generators, the \textit{Virasoro algebra}, which in turn completely determines (up to the so-called structure constants) the main quantities of interest, namely the correlation functions of certain special operators, thanks to a recursive procedure dubbed the \textit{conformal bootstrap}. 
 
A natural question which appeared shortly after these developments was: what happens when the algebra of symmetry \textit{stricly contains} the Virasoro algebra? In other words, do the same techniques apply when Virasoro symmetry is extended to feature an additional level of symmetry? Certain extensions of the Virasoro algebra, called $W$-algebras, have been first studied by Zamolodchikov in his work \cite{Za85} where was presented the notion of higher-spin symmetry, and following this work  two-dimensional CFTs having this extended level of symmetry appeared in the physics literature in \cite{FaLu,FaZa}. In addition to being an object of interest per se, the study of  CFTs with $W$-symmetry is crucial in the understanding of $W$-strings, $W$-gravity theories or certain statistical physics systems (the articles \cite{FaZa2} and \cite{JMO} provide explicit instances of such systems) and can be applied to the understanding of some Wess-Zumino-Novikov-Witten models (such a link is explained in \cite{FORTW}). From the representation theory viewpoint the study of W-algebras has proved to be a seminal topic with numerous applications ranging from integrable hierarchies to the geometric Langlands program (see \cite{Arakawa_intro} and the references therein).
 
Toda Conformal Field Theories are a family of two-dimensional CFTs indexed by (finite-dimensional) semisimple and complex Lie algebras $\mathfrak{g}$. One of their features is that they may be thought of as realizations of these algebras of symmetry since they are assumed to provide highest-weight representations of $W$-algebras. In this context, the primary fields, defined as random fields on a \green{two-dimensional} surface $\Sigma$ with special conformal covariance properties, are called Vertex Operators and  their correlations are defined as an average of the product of the fields (taken at different points of the surface $\Sigma$) with respect to the law of a random map called the \textit{Toda field}. This construction can be made somehow explicit thanks to the fundamental fact that Toda CFTs admit a path integral formulation: given a Riemannian metric $g$ on $\Sigma$ (with associated scalar curvature $R_g$, gradient $\partial_g$ and volume form ${\rm v}_g$), the Toda CFT associated to $\mathfrak{g}$ is a theory of fields $\phi:\Sigma\to \left(\mathfrak{a},\langle \cdot , \cdot \rangle\right)$ where $\left(\mathfrak{a},\langle \cdot , \cdot \rangle\right)$ is an Euclidean space naturally defined from $\mathfrak{g}$ (see Subsection \ref{Lie_reminders}). Within this framework, the (formal) path integral description of Toda CFT reads for arbitrary test functions $F$
\green{\begin{equation}\label{partition_function}
     \langle F \rangle_{T,g} \coloneqq \int F( \phi)e^{-S_{T,\mathfrak g}( \phi,g)}D \phi
 \end{equation}
where $D \phi$} refers to the putative ``uniform measure" on the space of square integrable \green{$\mathfrak{a}$}-valued maps defined on $\Sigma$ and $ S_{T,\mathfrak{g}}$ is the Toda action given by 
\green{\begin{equation}\label{action}
 S_{T,\mathfrak{g}}(\phi,g)\coloneqq  \frac{1}{4\pi} \int_{\Sigma}  \Big (  \langle\partial_g \phi(x), \partial_g \phi(x) \rangle_g   +R_g \langle Q, \phi(x) \rangle +4\pi \sum_{i=1}^{r} \mu_ie^{\gamma    \langle e_i,\phi(x) \rangle}   \Big)\,{\rm v}_{g}(dx),
\end{equation}}
with 
\begin{itemize}
\item
$(e_i)_{1\leq i\leq r}$ a special basis of $\mathfrak{a}$ made of so-called \emph{simple roots},
\item
$\langle \cdot , \cdot \rangle_g$ the scalar product associated to the tangent space of $\mathfrak{a}$-valued functions defined on $\Sigma$,
\item
the constants $\mu_i$ ($1\leq i\leq r$), positive and dubbed the cosmological constants,
\item
$\gamma>0$ the coupling constant,
\item
$Q$ the $\mathfrak{a}$-valued background charge.
\end{itemize}
In order to ensure conformal symmetry, the background charge is related to the coupling constant via the relation $Q\coloneqq \gamma\rho+\frac2\gamma\rho^\vee$ where $\rho$ and $\rho^\vee$ are special vectors in $\mathfrak{a}$ (see Subsection~\ref{Lie_reminders}. Let us emphasize  that one recovers Liouville CFT when $\mathfrak{g}$ is the Lie Algebra $\mathfrak{sl}_2$ (of $2\times 2$ complex matrices with vanishing trace, in which case $r=1$) and that the convention for $\gamma$ in this paper differs by a scaling factor of $\sqrt{2}$ from the standard convention for Liouville CFT in the probabilistic literature. The next section~\ref{section_background} is devoted to providing more details on these notations.

At this stage, it is worth mentioning that the action functional \eqref{action} has a relevant geometrical meaning in the context of \textit{$W$-geometry}, introduced by Gervais and Matsuo in \cite{GeMa92}. Though we will not prove the following claim in this paper, the connection with  \textit{$W$-geometry} can be recast in the probabilistic language as follows.
Up to renormalizing the measure \eqref{partition_function} by its total mass, the Toda field $\phi$ can be understood as a random  map from the Riemannian  surface $(\Sigma,g)$ to \green{$\mathfrak{a}$} whose law is described by  
 \begin{equation}\label{path_integral}
\forall \,F \text{ test function},  \quad   \E^g_{\gamma,\bm \mu}[F(\phi)]\coloneqq \frac{ \langle F \rangle_{T,g}}{\langle 1 \rangle_{T,g}}.
 \end{equation}
At this informal level of discussion, we expect that, in the semi-classical limit $\gamma\to0$ such that for all $1\leq i\leq r$,  $\mu_i= \frac{\Lambda}{\gamma^2}$ with fixed $\Lambda>0$, the law of the Toda field (rescaled by $\gamma$) will converge to the (classical) solution of the \textit{Toda equation} : 
\begin{equation}\label{Toda_equ}
    2 \Delta_g u = R_g\rho^\vee + 4 \pi \Lambda\sum_{i=1}^{r} e_ie^{\ps{e_i,u}}\quad\text{on }\Sigma.
\end{equation}
We expect this conjecture to be true at the level of large deviations (or semi-classical analysis) in the spirit of the paper \cite{LRV19}. Yet one has to be cautious here because the normalization constant  $\langle 1 \rangle_{T,g}$ may become infinite depending on the topology of $\Sigma$. In the \green{rest of the document}, we will focus on the case where $\Sigma$ is the Riemann sphere $\mathbb{S}^2$ (though the present framework can be extended to other topologies), in which case $\langle 1 \rangle_{T,g}=\infty$. Still one can define quantities with insertion of a certain number of Vertex Operators, whose parameters obey the \green{\textit{Seiberg bounds}}~\cite{Sei90} in order to ensure existence of the corresponding correlation function (see next subsection). Interestingly, it is explained in \cite{GeMa92} that in this setup \green{(that is $\Sigma=\S^2$)} $\mathfrak{sl}_n$ Toda equations (where one adds appropriate conical singularities) can in some sense be interpreted as compatibility equations for a meromorphic embedding of a two-sphere into a complex projective plane $\C\mathrm P^n$, \green{thus establishing} a correspondence between solutions of the Toda equation \eqref{Toda_equ} and certain meromorphic embeddings from $\C\mathrm{P}^1$ to $\C\mathrm{P}^n$, a problem which somehow provides a generalization of the celebrated \textit{uniformisation of Riemann surfaces}. More details can be found in Subsection~\ref{semi-classical}.

%Representation theoretical methods, \textit{i.e.} Vertex Operator Algebra \cite{Borcherds}, are developed mostly in the case of rational CFTs (roughly speaking, for CFTs with discrete spectrum). 
 \subsection{A probabilistic construction}
Though an algebraic approach to CFTs was developped shortly after the BPZ paper (see the notion of Vertex Operator Algebra \cite{Borcherds,FLM89}), a probabilistic approach to conformal invariance was only developed recently following the introduction by Schramm \cite{schramm} of random curves, called Schramm-Loewner Evolutions (SLEs), which describe (conjecturally at least) the interfaces of critical models of statistical physics (such as percolation or the Ising model). More recently, there has been a huge effort in probability theory to make sense of Liouville CFT within the realm of random conformal geometry and the scaling limit of random planar maps (see  \cite{LeG13,Mie13,GM20,DMS14,DDDF,DFGPS}). Another approach, based on the path integral formulation of Liouville CFT in the physics literature,  was developed in  \cite{DKRV,DRV16,HRV16,GRV16} to give a rigorous probabilistic construction of  the correlation functions of Liouville CFT. This construction initiated a program \cite{KRV_loc} to lay the mathematical  foundations of the conformal bootstrap procedure envisioned in physics by BPZ, namely that one can express the Liouville CFT correlation functions in terms of representation theoretical special functions. The building blocks are an 
  explicit formula for the three point correlation functions (or equivalently the structure constants) and a recursive procedure for the higher correlations. The explicit formula for the three-point structure constants discovered in the physics literature, the celebrated DOZZ formula, was recovered probabilistically in \cite{KRV_DOZZ} and a probabilistic justification of the conformal bootstrap formalism for the higher order correlations was provided  recently in \cite{GKRV}.

Building on \cite{DKRV}, our goal here is to provide a probabilistic definition of Toda CFTs in the  case where we consider the underlying Lie algebra to be a finite-dimensional simple and complex Lie algebra, and when the manifold on which the theory is constructed is the (Riemann) sphere. To do so we follow the ideas developed in the case of Liouville CFT (which corresponds to the $\mathfrak{sl}_2$ case) in~\cite{DKRV} and interpret the path integral formulation of the theory as a formal way of defining a measure on some functional space. More precisely we interpret the mapping defined via the path integral \eqref{partition_function} as a measure $F\mapsto\ps{F}_{T,g}$ on the Sobolev space with negative index $\mathrm{H}^{-1}(\S^2\to\green{\mathfrak{a}},g)$ (which we define in~\eqref{Hnorm}); in order to construct this measure we introduce a probabilistic framework which involves two objects: % that have become fundamental over the past few years
the Gaussian Free Field (GFF) and the exponential of the GFF called Gaussian Multiplicative Chaos (GMC). 
 
The presence of the GFF is related to the presence of the square gradient term in the Toda field action and has proven to be particularly relevant in the context of constructive conformal field theory. But as opposed to Liouville CFT where only one GFF is involved in the construction, in Toda CFTs we have to consider several GFFs that are coupled in a way that is prescribed by the underlying structure of the Lie algebra. The fields of Toda CFTs are well-defined but non-regular since they exist only in the sense of Schwartz distributions; therefore the exponential terms that appear in~\eqref{action} are not well-defined objects. However, GMC theory provides a way of making sense of these terms as (random) Radon measures.

This interpretation allows to construct a regularized partition function by taking ${F=1}$ in~\eqref{partition_function}; however and similarly to the existence of Seiberg bounds in Liouville CFT~\cite{Sei90} this partition function will not converge in relation with an obstruction of geometrical nature: the Gauss-Bonnet theorem entails that classical Toda equations (\textit{i.e} the equations of motion associated to this action) cannot admit solutions on the Riemann sphere. This difficulty can be overcome by looking at special functionals $F$ that admit Vertex Operators as factors; by adding these extra terms to the measure the partition function becomes the correlation function of Vertex Operators and is predicted to exist as long as some conditions on these operators are satisfied. To define these Vertex Operators one relies on a regularization of the   Toda field and introduces the regularized Vertex Operator $V_{\alpha,\eps}(z)$, which is expressed for $z$ on the Riemann sphere in terms of the Toda field $\green\Phi$ and a \textit{weight} $\alpha\in\green{\mathfrak a}$: up to constant terms, $V_{\alpha,\eps}(z)$ is defined as $\eps^{|\alpha|^2/2}e^{\langle \alpha, \green\Phi_\eps \rangle}$ where $\green\Phi_\eps$ is the field $\green\Phi$ smoothed up at scale $\eps$  (its definition will be made precise in Subsection \ref{vertex}). Our main result provides a necessary and sufficient condition that ensures the existence of the correlation functions (defined as limits of $\langle \prod_{k=1}^NV_{\alpha_k,\eps}(z_k)\rangle_{T,g} $ when the cut-off $\eps$ is sent to $0$). Moreover, the correlations are indeed conformally covariant as predicted by CFT: 
\begin{theorem}\label{construction}
Let $\mathfrak{g}$ be a finite-dimensional simple and complex Lie algebra and assume that $\gamma\in(0,\sqrt 2)$. If $g$ is any Riemannian metric in the conformal class of the standard round metric on the sphere $\hat g$, let $\langle \prod_{k=1}^NV_{\alpha_k,\eps}(z_k)\rangle_{T,g} $ be the regularized correlation function of the $\mathfrak{g}$-Toda theory. Then:
\begin{description}
\item[1. (Seiberg bounds)]  The limit
\begin{equation*}
\langle \prod_{k=1}^NV_{\alpha_k}(z_k)\rangle_{T,g}:= \underset{\eps \to 0}{\lim}  \: \langle \prod_{k=1}^NV_{\alpha_k,\eps}(z_k)\rangle_{T,g}
\end{equation*}
exists and is non trivial if and only if the two following conditions hold for all ${i=1,\dots,r}$: 
\begin{itemize}
\item
$\displaystyle\ps{\sum_{k=1}^N\alpha_k-2Q,\omega_i^\vee} >0$
where the $(\omega_i^\vee)_{1\leq i\leq r}$ are the fundamental coweights (\emph{i.e} form the basis dual to that of the simple roots),
\item
for all $1\leq k\leq N$, $ \langle\alpha_k-Q, e_i \rangle \:  <  \: 0$.
\end{itemize}
\item[2. (Conformal covariance)]  For any M\"obius transform of the plane $\psi$
 \begin{equation*}
 \langle \prod_{k=1}^NV_{\alpha_k}(\psi (z_k) )\rangle_{T,g}= \prod_{k=1}^N\norm{\psi'(z_k)}^{-2\Delta_{\alpha_k}}\langle \prod_{k=1}^NV_{\alpha_k}(z_k) \rangle_{T,g}.
 \end{equation*}
where the conformal weights are given by $\Delta_{\alpha_j}\coloneqq \ps{ \frac{\alpha_j}2, Q-\frac{\alpha_j}{2}}$.
\item[3. (Weyl anomaly)]  For appropriate $\varphi$ (more precisely $\varphi\in\bar{C}^1(\R^2)$: see notations in Section~\ref{section_background})  then  
 \begin{equation*}
  \langle \prod_{k=1}^NV_{\alpha_k}(z_k)\rangle_{T,e^\varphi \hat g} =e^{  \frac{\mathbf{c}_T}{96 \pi}S_L(\varphi,\hat g)  }  \langle \prod_{k=1}^NV_{\alpha_k}(z_k)\rangle_{T,\hat{g}}
 \end{equation*}
 where $S_L$ is the Liouville functional (with vanishing cosmological constant)
 $$S_L(\varphi,\hat g)\coloneqq \int_{\S^2 }\big(  |\partial_{\hat{g}} \varphi|^2_{\hat{g}}  + 2   R_{\hat{g}}   \varphi \big)  \,d{\rm v}_{\hat{g}}  ,$$ 
 and the central charge $\mathbf{c}_T$ is given by $\mathbf{c}_T= r+ 6 |Q|^2 $.
\end{description}
\end{theorem}
The value of the central charge can be described explicitly in terms of the coupling constant $\gamma$ and the underlying Lie algebra. Indeed, finite-dimensional and simple complex Lie algebras are completely classified and belong (up to isomorphism) to finitely many families of Lie algebras (see Subsection~\ref{Lie_reminders}) for which the central charge is explicit: see \eqref{Weyl_norm} or \eqref{central_charge}. 

Our main statement can be understood as a rigorous definition of the correlation functions of Toda CFTs, but also of the law of the   Toda field $\green\Phi$ (when we have fixed marked points $(\bm z, \bm \alpha)=(z_k,\alpha_k)_{1 \leq k \leq N}$ that satisfy the Seiberg bounds) by setting 
\begin{equation}
    \E^{(\bm z,\bm \alpha)}\left[F(\green\Phi)\right]\coloneqq \frac{\ps{F\prod_{i=1}^NV_{\alpha_i}(z_i)}_{T,g}}{\ps{\prod_{i=1}^NV_{\alpha_i}(z_i)}_{T,g}}
\end{equation}
where $F$ is any bounded and continuous functional on $\mathrm{H}^{-1}(\S^2\to\green{\mathfrak{a}},g)$ and $\ps{F\prod_{i=1}^NV_{\alpha_i}(z_i)}_{T,g}$ is the limit of $\langle F \prod_{k=1}^NV_{\alpha_k,\eps}(z_k)\rangle_{T,g} $ when the cut-off $\eps$ goes to $0$ (this more general case can be handled similarly to the case $F=1$). 

\begin{remark}
The above construction can be generalized when one considers as underlying Lie algebra any {\bf semisimple} and complex Lie algebra. Indeed, from the classification of semisimple Lie algebras (see for instance~\cite{Hum72}), such a Lie algebra can be written as a direct sum of \textbf{simple} Lie algebras $\mathfrak{g}=\oplus_{k=1}^p \mathfrak{g}_k$. Moreover a general property of Toda CFTs (which can be derived from the form of the Toda field action) is that for $A,B$ two semisimple and complex Lie algebras 
\begin{equation}
    \ps{\prod_{k=1}^NV_{(\alpha_k,\beta_k)}(z_k)}_{T, g}^{A\oplus B}=\ps{\prod_{k=1}^NV_{\alpha_k}(z_k)}_{T,g}^A\ps{\prod_{k=1}^NV_{\beta_k}(z_k)}_{T,g}^B,
\end{equation}
where with the notation $\ps{\cdot}_{T, g}^{\mathfrak{g}}$  we have stressed the dependence on the Lie Algebra $\mathfrak{g}$. This provides a way of constructing correlation functions for general finite-dimensional and semisimple complex Lie algebras. 
The latter equation also implies that the central charges add up, in the sense that
\begin{equation}
    \bm c_{T,A\oplus B}=\bm c_{T,A}+\bm c_{T,B}
\end{equation}
where here again the notation $ \bm c_{T,\mathfrak{g}}$ stresses the dependence on the  Lie Algebra $\mathfrak{g}$. 
\end{remark}

\subsection{\green{Future directions}}
The present document represents the starting point of a mathematical study of Toda CFTs, whose understanding has proved to be key in many domains of both the physics and mathematics community. In particular the probabilistic framework introduced to give a meaning to Toda CFTs will hopefully initiate a thorough investigation of this family of CFTs from a mathematical viewpoint. Let us highlight below some outlooks we find particularly relevant and that the setup introduced in this paper should hopefully allow to address.  

\subsubsection*{Integrability} One of the main specificities of Toda CFTs is the existence of a higher level of symmetry encoded by $W$-algebras. This symmetry arises via so-called \emph{Ward identities} involving holomorphic currents and that constrain correlation functions.
In the context of the $\mathfrak{sl}_3$ \green{Toda CFT}, it has been proved by Y. Huang and the first author in~\cite{Toda_OPEWV} that, thanks to the present probabilistic framework, it is possible to provide a rigorous meaning to such identities from which can be derived a \emph{BPZ-type} differential equation for certain four-point correlation functions. This should pave the way to the computation of some Toda three-point correlation functions, as predicted in the physics literature, and that generalize the celebrated DOZZ formula~\cite{DO94,ZZ96,KRV_DOZZ} for Liouville CFT. Subsection~\ref{sec:sym} provides additional details with respect to this integrability perspective.

\subsubsection*{Conformal bootstrap} The conformal bootstrap is a powerful tool for solving CFTs, based on a systematic exploitation of the constraints imposed by conformal symmetry. In Liouville CFT it takes the form of a recursive procedure to compute any correlation function from three-point correlation functions, given the knowledge of the so-called \emph{conformal blocks}. Recently it was shown by Guillarmou, Kupiainen and the last two authors in~\cite{GKRV} that such a procedure was indeed valid for Liouville CFT, based on a mathematically rigorous setup. Proving that such a machinery would work for Toda CFTs is more involved and at the moment remains an open question, even in the physics literature, and would require a spectral analysis of certain commuting Hamiltonians acting on a Fock space.  

\subsubsection*{W-geometry} In Liouville CFT, critical points of the action functional describe conformal metrics with constant negative curvature and represent classical fields of the theory. A standard way to recover such fields from the quantum model is to perform a semi-classical analysis, that is to study the behaviour of the theory when the coupling constant ---representing the level of randomness considered--- is taken to zero.  This has been successfully carried out by H. Lacoin and the last two authors in~\cite{LRV19} for Liouville CFT on the sphere. 
The question of performing a similar analysis and its meaning in the framework of Toda CFTs is discussed in Subsection~\ref{semi-classical} below.

\section{Background and notations}\label{section_background}
%%%%%%%%%%%%%%%%%%  
%We consider on the  space $\R^2$ its usual inner  product $(\cdot,\cdot)$. %We also introduce the space of smooth functions that are compactly supported in $\R^2$, which we denote as $C^\infty_c(\R^2)$. 

\subsection{Some reminders on conformal geometry and Lie algebras}
 %%%%%%%%%%%%%%%%%%%

\subsubsection{Conformal geometry on the Riemann sphere}

The  sphere $\S^2$ can be mapped by stereographic projection to the (compactified) plane (\textit{i.e.} the Riemann sphere) which we view both as ${\R^2}\cup\lbrace\infty\rbrace$ and $\C\cup\lbrace\infty\rbrace$. We will work under this more convenient framework in the sequel.

\medskip
{\bf Metrics on the Riemann sphere.}

 We will consider  differentiable conformal metrics on the two-dimensional sphere $\S^2$; they can be identified via stereographic projection with metrics on the plane of the form $g=e^{\varphi}\hat{g}$ with $\hat{g}$ is the standard round metric 
  \begin{equation}
    \hat{g}\coloneqq \frac{4}{(1+|x|^2)^2}|dx|^2,
\end{equation}%=\frac{2}{(1+\bar zz)^2}(dz\otimes d\bar z+d\bar z\otimes dz)$$
and $\varphi\in\bar{C}^1(\R^2)$ where, for $k\geq 0$, $ \bar{C}^k(\R^2)$ stands for the space of functions $\varphi:\R^2\to\R$ for which both $\varphi$ and $x\mapsto\varphi (1/x)$ are $k$-times differentiable with continuous derivatives.   
The reader may check that the metric $ \hat{g}$ is the pushforward (via stereographic projection) of the standard metric on the Riemann sphere $\S^2$.  We will thus work with such metrics $g$ on the plane, for which we will denote by $\partial_g$ the gradient, $\triangle_g$ the Laplace-Beltrami operator, $R_g=-\triangle_g\ln \sqrt{\det g}$ the Ricci scalar curvature and ${\rm v}_g$ the volume form. If $u,v\in\R^2$, we denote by $(u,v)_{g}$ the inner product with respect to the metric $g$ ($|\cdot |_{g}$ stands for the associated norm). When no index is given, this means that the object has to be understood in terms of the usual Euclidean metric on the plane (\textit{i.e.} $\partial$, $\triangle$, $R$, ${\rm v}$ and $(\cdot,\cdot)$).  
Since the stereographic projection is an isometry, we already know that the spherical metric $\hat g$ is such that $R_{\hat g}=2$ (its Gaussian curvature is $1$) with total mass ${\rm v}_{\hat g}(\R^2)=4\pi$.

More generally, two metrics $g$ and $g'$ will be said to be conformally equivalent when $$g =e^{\varphi}g'$$ for  $\varphi \in \bar{C}^1(\R^2) $. It is readily seen that as soon as $g'$ is in the conformal class of the spherical metric ---that is when $g'=e^{\varphi}\hat g$ with $\varphi \in \bar{C}^1(\R^2) $--- one has ${\int_{\R^2 }\big(  |\partial_{g'} \varphi|^2_{g'}  + 2   R_{g'}   \varphi \big)  \,d{\rm v}_{g'} <\infty}$. Furthermore, for $\varphi \in \bar{C}^2(\R^2) $,  the curvatures of two such metrics are related by the relation
\begin{equation}\label{curvature}
R_g=e^{-\varphi}\big(R_{g'}-\Delta_{g'}\varphi\big).
\end{equation}

In what follows and for given metrics $g$ and $h\in \bar C^1(\R^2)$, we will denote by $m_{g}(h)$ the mean value of $h$ in the metric $g$, that is the quantity
\begin{equation}
m_{g}(h)\coloneqq \frac{1}{{\rm v}_g(\R^2)}\int_{\R^2}h(x)\,{\rm v}_g (dx)
\end{equation}
and work in the Sobolev space $\mathrm H^1(\R^2,g)$, which is the closure of $C^\infty_c(\R^2)$ with respect to the Hilbert-norm
\begin{equation}\label{Hnorm}
\int_{\R^2}h(x)^2\,{\rm v}_{g}(dx)+\int_{\R^2}|\partial_g h(x)|^2_g\,{\rm v}_g(dx).
\end{equation}
The continuous dual of $\mathrm H^1(\R^2,g)$ will be denoted $\mathrm H^{-1}(\R^2,g)$. It may be useful to note that the Dirichlet energy is a conformal invariant, that is to say is independent of the metric within a given conformal class:
\begin{equation}\label{diri}
\int_{\R^2}|\partial_{g'} h(x)|_{g'}^2\, {\rm v}_{g'}(dx)=\int_{\R^2}|\partial_{g} h(x)|_{g}^2\,{\rm v}_{g}(dx).
\end{equation}

\medskip
{\bf Green kernels.} Given a metric $g$ on the Riemann sphere that is conformally equivalent to the spherical metric $\hat g$, we denote by $G_g$ the Green function of the problem
\begin{equation}\nonumber
\triangle_g u=-2\pi \left(f-m_g(f)\right) \quad\text{on }\R^2,\quad  \int_{\R^2} u(x) \,{\rm v}_g(dx) =0
\end{equation}
where $f$ belongs to the space $L^2(\R^2, g)$ and $u$ is in $ \mathrm H^1(\R^2,g)$.
This means that the solution $u$ can be expressed as
\begin{equation}
u=\int_{\R^2} G_g(\cdot,x)f(x){\rm v}_g(dx)=:G_gf
\end{equation}
with $m_g(G_g(x,\cdot))=0$ for all $x\in\R^2$.
The kernel $G_g$ has an explicit expression given by (see \cite[Equation (2.9)]{DKRV}) 
\begin{equation}
    G_g(x,y)=\ln\frac{1}{\norm{x-y}}-m_g\left(\ln\frac{1}{\norm{x-\cdot}}\right)-m_g\left(\ln\frac{1}{\norm{y-\cdot}}\right)+\theta_g
\end{equation}
where \[\theta_g\coloneqq \frac{1}{{\rm v}_g(\R^2)^2}\int_{\R^2}\int_{\mathbb{R}^2}\ln\frac{1}{\norm{x-y}}{\rm v}_g(dx){\rm v}_g(dy).
\]
For instance for the spherical metric this becomes
\begin{equation}\label{Green_round}
    G_{\hat g}(x,y)=\ln\frac{1}{\norm{x-y}}-\frac14\left(\ln\hat g(x)+\ln\hat g(y)\right)+\ln 2-\frac12.
\end{equation}

Another well-known property of these Green functions (see \cite[Proposition 2.2]{DKRV} for instance) is that they are conformally covariant in the sense that: 
\begin{lemma}[Conformal covariance]\label{conformal_covariance_green}
Let $\psi$ be a M\"obius transform of the Riemann sphere and $g$ be a Riemannian metric conformally equivalent to the spherical one. Then
\begin{equation}
    G_{g_{\psi}}(x,y)=G_g(\psi(x),\psi(y))
\end{equation}
where $g_{\psi}(z)=\norm{\psi'(z)}^2g(\psi(z))$ is the pullback of the metric $g$ by $\psi$.
\end{lemma}
Again let us register what happens for the spherical metric:
\begin{equation}\label{conformal_spherical}
    G_{\hat g}(\psi(x),\psi(y))=G_{\hat g}(x,y)-\frac14(\phi(x)+\phi(y))
\end{equation}
where $\phi$ is such that $e^{\phi}=\frac{\hat g_{\psi}}{\hat g}$.

\subsubsection{Lie algebras and the Toda field action}\label{Lie_reminders}
We provide here the background on Lie algebras needed to make sense of the path integral definition of Toda CFTs. We will be very synthetic and refer for instance to the textbook~\cite{Hum72} for additional details.

{\bf Finite-dimensional simple and complex Lie algebras.} 
Simple Lie algebras are (non-Abelian) Lie algebras enjoying the remarkable property that they do not admit any proper, nonzero ideals. When they are finite-dimensional and complex, such an assumptions leads to a classification of such Lie algebras up to isomorphism. Namely a simple and complex Lie algebra is either isomorphic to a classical Lie algebra, that is one of the Lie algebras $(A_n)_{n\geq 1}$ (corresponding to $\mathfrak{sl}_{n+1}$), $(B_n)_{n\geq 2}$ (for $\mathfrak{o}_{2n+1}$), $(C_n)_{n\geq 3}$ ($\mathfrak{sp}_n$) and $(D_n)_{n\geq 4}$ ($\mathfrak{o}_{2n}$), or an exceptional Lie algebra, that is either $E_6$, $E_7$, $E_8$, $F_4$ or $G_2$.

To any finite-dimensional simple and complex Lie algebra $\mathfrak{g}$ is naturally attached an Euclidean space $(\mathfrak{a},\ps{\cdot,\cdot})$. This finite-dimensional real vector space is such that the Cartan subalgebra of $\mathfrak{g}$ can be written as $\mathfrak{a}\oplus i\mathfrak{a}$, and comes equipped with a (positive definite) scalar product $\ps{\cdot,\cdot}$. This scalar product is inherited from the Killing form $\kappa$ of $\mathfrak{g}$ in that both are proportional one to the other. This Euclidean space is unique up to isomorphism and can be thought of as $\R^r$ endowed with its standard scalar product, where $r$ is the \emph{rank} of $\mathfrak{g}$. This Euclidean space also comes with a special basis $(e_i)_{1\leq i\leq r}$ made of so-called \emph{simple roots}. This basis satisfies the property that
\begin{equation}
    2\frac{\ps{e_i,e_j}}{\ps{e_i,e_i}}=A_{i,j} \quad\text{for all }1\leq i,j\leq r
\end{equation}
where $A$ is the Cartan matrix of $\mathfrak{g}$, explicit for $\mathfrak{g}$ as considered.
For instance the $\mathfrak{sl}_n$ Cartan matrix is  tridiagonal with $2$ on the diagonal and $-1$ on the entries $(i,j)$ with $|i-j|=1$. In general the entries of this matrix are integral, equal to $2$ on the diagonal and non-positive elsewhere; the matrix is invertible. It is common to renormalize the scalar product so that the longest roots have squared norm $2$, which we will do in the sequel. The renormalization constant used %is usually referred to as the \textit{Dynkin index of the adjoint representation} and 
is given by $2h^\vee$, where $h^\vee$ is the so-called \textit{dual Coxeter number}, an explicit positive integer that depends on the underlying Lie algebra.

It is very natural to introduce the basis of the \textit{fundamental weights} $(\omega_i)_{1\leq i \leq r}$, which is a basis of $\mathfrak{a}$ defined by setting
 \begin{equation}
 \omega_i\coloneqq  \sum_{l=1}^{r} (A^{-1})_{i,l}e_l.
 \end{equation} 
They are defined so that  ($ \delta_{ij}$ is the Kronecker symbol)
 \begin{equation}\label{relweights}
\langle e_i^{\vee},\omega_j\rangle= \delta_{ij}  ,\quad  \langle \omega_i,\omega_j \rangle=  \sum_{l,l'=1}^{r} (A^{-1})_{i,l} A_{l,l'} (A^{-1})_{l',j}=(A^{-1})_{i,j}
 \end{equation}
 where $e_i^{\vee}\coloneqq 2\frac{e_i}{\ps{e_i,e_i}}$ is the coroot associated to $e_i$. The \textit{Weyl vector} which is defined by
 \begin{equation}
     \rho\coloneqq \sum_{i=1}^r \omega_i
 \end{equation}
 naturally enjoys the property that $\ps{\rho,e_i^{\vee}}=1$ for all $1\leq i\leq r$. We will also consider the Weyl vector associated to the coroots by considering the vector $\rho^\vee=\sum_{i=1}^r\omega_i^\vee$ where the $(\omega_i^\vee)_{1\leq i\leq r}$ are defined in such a way that  $\ps{\omega_i^{\vee},e_j}=\delta_{i,j}$ for all $1\leq i,j\leq r$.
  The squared norm of the Weyl vector can be expressed explicitly \green{in terms of} the Lie algebra under consideration via the \textit{Freudenthal-de Vries strange formula} for simple Lie algebras \cite[Equation (47.11)]{FdV}\footnote{This equation differs from the one in \cite{FdV} by a multiplicative factor $2h^\vee$. This is due to our normalization convention for the scalar product $\ps{\cdot,\cdot}$ on $\mathfrak{a}^*$.}
 \begin{equation}
 \norm{\rho}^2=\frac{h^\vee\dim \mathfrak{g}}{12}\cdot
 \end{equation}
Using the explicit values of $h^\vee$ and $\dim \mathfrak{g}$ this quantity is seen to given by
\begin{equation}\label{Weyl_norm}
\begin{split}
\frac{n(n+1)(n+2)}{12}\text{ for }A_n,\quad\frac{n(2n-1)(2n+1)}{12}\text{ for }B_n,\\
\frac{n(n+1)(2n+1)}{12}\text{ for }C_n,\quad\frac{(n-1)n(2n-1)}{6}\text{ for }D_n,
\end{split}
\end{equation}
and $78$, $\frac{399}{2}$, $620$, $39$, $\frac{14}{3}$ for the exceptional Lie algebras $E_6$, $E_7$, $E_8$, $F_4$ and $G_2$.
More generally we can explicitly compute the values of $\norm{\rho^\vee}^2$ and $\ps{\rho,\rho^\vee}$ in all the cases considered (see Table~\eqref{central_charge} below).

To finish this quick introductory part on simple Lie algebras let us mention the following duality relation between two vectors $\alpha,\beta\in\mathfrak{a}$
 \begin{equation}\label{dual}
 \langle\alpha, \beta \rangle= \sum_{i=1}^{r}  \langle\alpha,\omega_i \rangle  \langle\beta,e_i^{\vee} \rangle,
 \end{equation}
which follows from the fact that, since $\ps{\omega_i,e_j^\vee}=\delta_{ij}$, one has $\alpha=\sum_{i=1}^r\ps{\alpha,\omega_i}e_i^\vee$.
 
\medskip 
{\bf Toda field action.} Given $\varphi$ and $\phi$ two differentiable maps from $\R^2$ to $\mathfrak a$\footnote{The scalar field $\varphi$ being studied in Toda CFTs usually has values in $\mathfrak{a}^*$. To keep notations simple we adopt the convention that $\varphi$ actually takes values in the space of roots $\mathfrak{a}$. This identification is possible thanks to the Riesz representation theorem.}, $\varphi=\sum_{i=1}^{r}\varphi_i \omega_i$ and $\phi=\sum_{i=1}^{r}\phi_i\omega_i$, let us set
 \begin{equation}\label{defscalar}
 \langle\partial_g\varphi ,\partial_g\phi  \rangle_g\coloneqq \sum_{i,j=1}^{r} \langle \omega_i,\omega_j \rangle(\partial_g\varphi_i,\partial_g\phi_j)_g. 
\end{equation}
Recall that we have defined the Toda field action $S_{T,\mathfrak{g}}$ in the metric $g$ for the Lie algebra $\mathfrak{g}$ by the expression
 \begin{equation}
 S_{T,\mathfrak{g}}(\phi,g)\coloneqq  \frac{1}{4\pi} \int_{\R^2}  \Big (  \langle\partial_{g} \phi(x), \partial_{g} \phi(x) \rangle_{g}   +R_g \langle Q, \phi(x) \rangle +4\pi \sum_{i=1}^{r} \mu_i e^{\gamma    \langle e_i,\green\phi(x) \rangle}   \Big)\,{\rm v}_{g}(dx)
 \end{equation}
where we have introduced the background charge
 \begin{equation}
 Q\coloneqq \gamma\rho+\frac2\gamma\rho^\vee,
\end{equation}
$\bm \mu\coloneqq (\mu_1>0,\cdots,\mu_r>0)$ are the cosmological constants and $\gamma >0$ is the coupling constant. In the sequel, we assume that $\gamma$ satisfies the condition
 \begin{equation}\label{condgamma}
 \gamma \in (0,\sqrt{2}).
 \end{equation}
 The condition on $\gamma$ is the optimal condition ensuring that the probabilistic construction makes sense\footnote{Recall that here our convention on $\gamma$ is different from the standard convention for Liouville CFT in the probabilistic literature by a factor of $\sqrt{2}$.}: this will become clear later when connecting to GMC theory. For later purpose we stress the following crucial property of the background charge $Q$:
 \begin{equation}\label{eq:propQ}
  \text{For all } 1 \leq i \leq r,\quad \langle Q, e_i \rangle =\gamma\frac{\ps{e_i,e_i}}{2}+\frac{2}{\gamma}\quad\text{and}\quad\langle Q, e_i^\vee \rangle =\gamma+\frac{4}{\gamma\ps{e_i,e_i}}\cdot
\end{equation}
 The path integral we aim to construct corresponds to a measure on a suitable space of maps $\phi: \R^2\to  \green{\mathfrak{a}}$ formally corresponding to 
\begin{equation}\label{TQFTPI}
e^{- S_{T,\mathfrak{g}}(\phi,g)}D\phi.
\end{equation}
As we will explain below, this can be achieved thanks to the introduction of the GFF.

\subsection{Probabilistic interpretation of the path integral}
 %%%%%%%%%%%%%%%%%%%
 
 \subsubsection{Gaussian measure interpretation of the squared gradient}
 %%%%%%%%%%%%%%%%%%%%%%%%%%%%%
The Toda field action can be decomposed as a sum of different terms. The first one is the quadratic term 
\green{\[
    \frac{1}{2\pi}\int_{\R^2}\ps{\partial_g\phi(x),\partial_g\phi(x)}_g\mathrm v_g(dx)=\frac{1}{2\pi}\int_{\R^2}\ps{\phi(x),-\triangle_g\phi(x)}_gv_g(dx)=:\ps{\phi,-\triangle_g\phi}_g
\]} which is reminiscent of a Gaussian measure. Indeed, the measure formally written as 
\green{\begin{equation}\label{GPI}
e^{-\frac{1}{2}\ps{\phi,-\triangle_g\phi}_g}D\phi,
\end{equation}}
when restricted to the space $$\Sigma\coloneqq \{\green\phi\in \mathrm H^{-1}(\R^2\to\green{\mathfrak{a}},g); \int_{\R^2} \green\phi(x)\, {\rm v}_g(dx)=0\}$$ where $ \mathrm H^{-1}(\R^2\to\green{\mathfrak{a}},g)$ is the set of $\green{\mathfrak{a}}$-valued (generalized) functions with each component (with respect to a basis of $\green{\mathfrak{a}}$) in $\mathrm{H}^{-1}(\R^2,g)$, can be understood as the measure on a Gaussian space $\left(\ps{\green\phi,-\triangle_g f}_g\right)_{f\in \mathrm H^{1}}$ with covariance kernel given by $\ps{h,-\triangle_g f}_g$. In other words we are looking for a Gaussian field enjoying the property that
\begin{equation}\label{eq:GFF_cov1}
\expect{\ps{\green\phi,-\triangle_g f}_g\ps{\green\phi,-\triangle_g h}_g}=\ps{h,-\triangle_g f}_g
\end{equation}
for $f,g\in \mathrm H^{1}(\R^2\to\green{\mathfrak{a}},g)$. When $r=1$, this is achieved by introducing the GFF \green{$X^g$} with vanishing ${\rm v}_g$-mean on the sphere, that is a centered Gaussian random distribution with covariance kernel given by the Green function $G_g$ (see \cite{dubedat,She07} for more details on this object). An important feature of the GFF is that it is not defined pointwise but rather belongs to the distributional space $\mathrm H^{-1}(\R^2,g)$. For generic rank this is done by considering additional fields and setting
 \begin{equation}
X^g\coloneqq \sum_{i=1}^{r}X^g_i\omega_i^\vee,
\end{equation}
where $X^g_1,\dots, X^g_{r}$ are $r$ such GFFs with covariance structure given by
\begin{equation}\label{GFF_covariance}
 \E[X^g_i(x) X^g_j(y)]=\ps{e_i,e_j}   G_g(x,y) .
 \end{equation}
This property implies that for any pair of vectors $u,v\in\mathfrak{a}$ we have
\begin{equation}\label{GFF_covariance2}
 \E[\ps{X^g(x),u}\ps{X^g(y),v}]=\ps{u,v}   G_g(x,y) .
 \end{equation}
 
To summarize we may wish to interpret the formal Gaussian measure \eqref{GPI} restricted to $\Sigma$ as 
\begin{equation*}\label{def:GPIr}
Z(g)^{-1}\green{\int F(\phi)e^{-\frac{1}{4\pi}\int_{\R^2}  \langle\partial_{g} \phi(x), \partial_{g} \phi(x) \rangle_{g}\,{\rm v}_{g}(dx) }D\phi}=\E\Big[F(X^g)\Big]
\end{equation*}
for \green{any} continuous and bounded functional $F$ on $\mathrm H^{-1}(\R^2\to \green{\mathfrak{a}},g)$, where $Z(g)$ stands for the total mass of the Gaussian integral \eqref{GPI}
\begin{equation*}
\Z(g)\coloneqq \left(\frac{\det (\bm\Sigma)}{\text{vol}_g(\R^2)^r}\right)^{-\frac12},
\end{equation*}
where $ \bm\Sigma$ is the covariance matrix
\begin{equation}
    \bm\Sigma\coloneqq\frac{-\triangle_{g}}{2\pi}\otimes A
\end{equation}
with $A$ the Cartan matrix of $\mathfrak g$ and $ \det (\bm\Sigma)$ is given by a regularized determinant. To be more specific, the Laplacian $\frac{-\Delta_g}{2\pi}$ acting over $\mathrm H^{1}(\R^2\to\R,g)$ has positive (apart from the zero eigenvalue) and discrete spectrum $(\lambda_j)_{j\geq 0}$, thanks to which one can define its spectral Zeta function $\zeta(s)\coloneqq \sum_{j\geq 1}\lambda_j^{-s}$ for $\mathfrak{Re}(s)\gg1$. The associated regularized determinant is then given by
\[
\det(\frac{-\Delta}{2\pi})=\exp\left(-\partial_{s}\zeta(s)_{\vert s=0}\right).
\]
A remarkable property of the partition function $Z(g)$ is its variation under a conformal change of metric. Indeed in the $r=1$ case it is proved in~\cite[Equation (1.13)]{OPS88} that
\begin{equation*}
    \log \frac{\det(-\triangle_{g'})}{\text{vol}_{g'}(\R^2)^r}=\log \frac{\det(-\triangle_{g})}{\text{vol}_g(\R^2)^r}+\frac{1}{96\pi}\int_{\R^2}\left(\norm{\dd \varphi}^2_g+2R_g\varphi\right){\rm v}_{g}(dx)
\end{equation*}
with $g'=e^{\varphi}g$. For $r>1$ since $\triangle$ acts independently on the $r$ components of a map $\R^2\to\R^r$ we see that
$\det(-\triangle_{g})=\det(-\triangle^1_{g})^r$ where $\triangle^1$ denote the Laplace operator acting on $\mathrm H^{1}(\R^2\to\R,g)$. Therefore
\begin{equation}
    \log \frac{\det(-\triangle_{g'})}{\text{vol}_{g'}(\R^2)^r}=\log \frac{\det(-\triangle_{g})}{\text{vol}_g(\R^2)^r}+\frac{r}{96\pi}\int_{\R^2}\left(\norm{\dd \varphi}^2_g+2R_g\varphi\right){\rm v}_{g}(dx).
\end{equation}

As a consequence, up to a global factor, one has  
\begin{equation}\label{reg_det}
    Z(e^{\varphi}\hat g)=\det(A)^{-\frac12}e^{\frac{r}{96\pi}\int_{\R^2}\left(\norm{\dd \varphi}^2_{\hat g}+2R_{\hat g}\varphi\right){\rm v}_{g}(dx)}
\end{equation}
within the conformal class of the spherical metric $\hat g$.

However in the above construction we do not take into account the fact that the GFF \green{$X^g$} has zero mean in the metric $g$ and, because the kernel of the Laplace operator $\triangle$ over $\mathrm H^{1}(\R^2\to\green{\mathfrak{a}},g)$ consists of constant functions, we can actually shift the field $X^g$ by a constant function without affecting Equation~\eqref{eq:GFF_cov1}. To overcome this issue we will introduce so-called\textit{ zero modes} in the interpretation of the squared gradient term as a Gaussian measure.
To do so we introduce the Lebesgue measure $d\bm c$ on $\green{\mathfrak{a}}$ by setting for each positive measurable function $F:\green{\mathfrak{a}}\to\R$
 \begin{equation}
 \int_{\green{\mathfrak{a}}}F(\bm c)\,d\bm c\coloneqq \det(A)^{\frac12}\int_{\R^{r}}F\Big(\sum_{i=1}^{r}c_i\omega_i^\vee\Big)\,dc_1\dots dc_{r},
\end{equation}
 where $dc_1,\dots, dc_{r}$ stands for the Lebesgue measure with respect to each variable $c_i$\footnote{The prefactor $\det(A)^{\frac12}$ in the equation  comes from the fact that the basis $(\omega_i^\vee)_{1 \leq i \leq r}$ is not orthonormal.}. 
% Here the Lebesgue measure has to be thought of as a Gaussian measure but whose variance is infinite. 
 By doing so we are led to the following probabilistic interpretation of the formal (full) Gaussian measure \eqref{GPI}
 \begin{equation}\label{def:GPI}
\green{\int F(\phi)e^{-\frac{1}{2}\langle \phi, -\triangle_{g} \phi \rangle_{g} }D\phi
}=Z(g) \int_{\green{\mathfrak{a}}}\E\Big[F\Big( X^g+\bm c)\Big)\Big]\,d\bm c 
\end{equation}
for each continuous and bounded functional $F$ on $\mathrm H^{-1}(\R^2\to \green{\mathfrak{a}},g)$ and $g$ in the conformal class of the spherical metric.
  \subsubsection{Gaussian Multiplicative Chaos interpretation of the exponential potential}
 %%%%%%%%%%%%%%%%%%%%%%%%%
 
It remains to treat the other terms that appear in the Toda field action \eqref{action}:
\[
\frac{1}{4\pi} \int_{\R^2}  \Big (R_g(x) \langle Q, \green\phi(x) \rangle +4\pi \sum_{i=1}^{r} \mu_i e^{\gamma    \langle e_i,\green\phi(x) \rangle}   \Big)\,{\rm v}_{g}(dx).
\]
The first term perfectly makes sense if we remember that the GFF has a meaning in the distributional sense. However the second term does not make sense because of the lack of regularity of the field. For it to be meaningful we need to make use of the notion of \textit{Gaussian Multiplicative Chaos} (see \cite{Kah,review}) which relies on a proper renormalization of some regularization of the GFF.

\begin{definition}
Let $\eta_\eps\coloneqq \frac1{\eps^2}\eta(\frac{\cdot}{\eps})$ be a smooth mollifier. We define the regularized field $X_\eps$ by considering the convolution approximation of $X$:
\begin{equation}
    X_\eps\coloneqq X*\eta_\eps.
\end{equation}
\end{definition}
\begin{definition}
Assume that $\gamma<\sqrt 2$ and consider the GFF $X^g_i=\langle e_i,  X^{g} \rangle$ from Equation~\eqref{GFF_covariance}. From  \eqref{covX} and basics of GMC theory \cite{review}, the following convergence holds in probability in the space of Radon measures (equipped with the weak topology): 
\begin{equation}
    M^{\gamma e_i, g}(dx)\coloneqq \underset{\eps \to 0}{\lim}  \:  e^{  \langle\gamma e_i,  X^{g}_\eps(x) \rangle-\frac12\expect{\langle\gamma e_i,  X^{g}_\eps(x) \rangle^2}} {\rm v}_{g}(dx).
\end{equation}
The random measure $M^{\gamma e_i,g}(dx) $ is non trivial (\textit{i.e.} different from $0$) and is called the Gaussian Multiplicative Chaos (GMC) measure associated to the field $\langle\gamma e_i,  X^{g} \rangle$.
\end{definition}
More generally the GMC measure
\begin{equation*}
M^{\alpha,g}(dx)\coloneqq  \underset{\eps \to 0}{\lim}  \:  \green{e^{  \langle\alpha,  X^{g}_\eps(x) \rangle-\frac12\expect{\langle\alpha,  X^{g}_\eps(x) \rangle^2}} {\rm v}_g(dx)}
 \end{equation*} 
exists and is non trivial if and only if $\green{\ps{\alpha,\alpha}< 4}$. 

\begin{remark}
The statement of \cite[Proposition (2.5)]{DKRV} can be easily adapted to show that the regularized GFF thus defined has a variance which evolves as
\[
   \expect{X^{\hat g}_{i, \eps}(x)X^{\hat g}_{j, \eps}(x)}=\ps{e_i,e_j}\left(-\ln\eps-\frac12\ln\hat g(x)+\theta_{\eta}+o(1)\right)
\]
when $\eps$ goes to $0$, and where we have set $\theta_{\eta}\coloneqq \int_\C\int_\C\eta(x)\eta(y)\ln\frac{1}{\norm{x-y}}\mathrm v(dx)\mathrm v(dy)+\ln 2-\frac12$ (recall that $\mathrm v$ denotes the standard Lebesgue measure). As a consequence the GMC measure defined above and the limiting measure defined by 
\[
 \underset{\eps \to 0}{\lim}  \:  \eps^{\gamma^2}e^{ \langle\gamma e_i,  X^{\hat g}_\eps(x) +\frac Q2\ln\hat g-\frac{\gamma e_i\theta_{\eta}}2\rangle }\mathrm v(dx)
\]
actually define the same random measure.
\end{remark}
 
In the end we interpret the path integral of Toda CFTs in a probabilistic way by making the identification for $F \in \mathrm H^{-1}(\R^2\to\green{\mathfrak{a}},g)$:
 \begin{multline}\label{def:TPI}
Z(g)^{-1} \green{\int_\Sigma F(\phi)e^{-S_{T,\mathfrak{g}}(\phi,g) }D\phi}\\
:=\lim\limits_{\eps\rightarrow0}\int_{\green{\mathfrak{a}}}\E\Big[F\Big( X^g_\eps+\frac Q2\ln g+\bm c)\Big)e^{-\frac{1}{4 \pi} \int_{\mathbb{R}^2} R_g(x)  \langle Q, X^g_\eps(x)+\bm c \rangle {\rm v}_ g(dx)  -\sum_{i=1}^{r} \mu_i e^{\gamma  \langle e_i, \bm c\rangle }M^{ \gamma e_i,g}_\eps(\C) }\Big]\,d\bm c,
\end{multline}
when the limit exists and where $g=e^{\varphi}\hat g$ is in the conformal class of the spherical metric.

\subsubsection{Vertex Operators}\label{vertex}
 There is a class of functionals $F$ which play a key role in the study of Toda CFTs; usually referred to as \textit{Vertex Operators}, computing their correlation functions is often one of the main issue in the study of two-dimensional CFTs. As we will see below, they enjoy a certain conformal covariance identity which is the starting point in the understanding of the theory. 
 
In Toda CFTs these Vertex Operators formally correspond to taking $F\green{(\phi)}=e^{\ps{\alpha,\green\phi(z)}}$ for $z\in\R^2$ and $\alpha\in\green{\mathfrak{a}}$ but since such $F$ are not defined on $\mathrm H^{-1}(\R^2\to\green{\mathfrak{a}},g)$, one must use a regularization procedure in order to define their correlations. This motivates the following definition:
 \begin{definition}
 For $z\in \C$ and $\alpha\in  \green{\mathfrak{a}}$ the regularized Vertex Operator $V^g_{\alpha,\eps}(z)$ is defined by 
  \begin{equation}
 V^g_{\alpha,\eps}(z)\coloneqq  \eps^{ \frac{|\alpha|^2}{2}}  e^{  \langle\alpha, X^g_\eps(z) + \frac Q2\ln g+\bm c-\frac{\alpha\theta_{\eta}}2\rangle}
 \end{equation}
  where $X^g_\eps(z)$ is the field regularized as above.
\end{definition}
Similarly to the GMC measure, this regularized Vertex Operator has same limit when $\eps\rightarrow0$ as the Wick exponential
\begin{equation}\label{VO_wick}
    e^{  \langle\alpha,  X^{\hat g}_\eps(x) + \bm c\rangle-\frac12\expect{\langle\alpha,  X^{\hat g}_\eps(x) \rangle^2}} \hat g(x)^{\ps{\frac{\alpha}2,Q-\frac{\alpha}2}}.
\end{equation}
Indeed for $\alpha\in \green{\mathfrak{a}}$ the variable $ \langle\alpha, X^g_\eps \rangle$ is given by the following formula
 \begin{equation*}
  \langle\alpha, X^g_\eps \rangle= \sum_{i=1}^{r} \langle\alpha,\omega_i\rangle X^g_{i,\eps}
 \end{equation*} 
 and thus has the following covariance structure, for $\alpha,\beta\in\green{\mathfrak{a}}$  
 \begin{equation}\label{covX}
 \E[     \langle\alpha, X^g_\eps(x) \rangle   \langle\beta, X^g_\eps(y) \rangle     ] =    \langle\alpha, \beta  \rangle  G_{g,\eps}(x,y)
 \end{equation} 
 where $G_{g,\eps}$ is the covariance kernel of the field $X^g_{\eps}$.
    
%  \textbf{In what follows if $g=\hat{g}$ is the round metric, we will omit the subscript (or superscript) in all notations: for example, $X^g$ will be denoted $X$, $G_{\hat{g}}$ will be denoted $G$, etc...}  

  %%%%%%%%%%%%%%%%%%%%%%%%%%%%%%%%%%%%%%%%%%%%%
\section{General definitions and existence theorems}
%%%%%%%%%%%%%%%%%%%%%%%%%%%%%%%%%%%%%%%%%%%%%
Having introduced the probabilistic tools allowing to translate in mathematical terms the path integral formulation of Toda CFTs, we are now ready to investigate the existence of the partition function and of the correlation functions.

 \subsection{\green{Toda CFT} measure and correlation functions}
 %%%%%%%%%%%%%%%%%
 
 To start with, recall that we have given in \eqref{def:TPI} a meaning to the measure on the space $\mathrm H^{-1}(\R^2\to \green{\mathfrak{a}},g)$ formally defined by the expression \eqref{TQFTPI} by setting for $F$ a positive measurable function on $\mathrm H^{-1}(\R^2\to \green{\mathfrak{a}},g)$:
\begin{multline*}
\langle F\rangle_{T,g}
:=Z(g)\lim\limits_{\eps\rightarrow0}\int_{\green{\mathfrak{a}}}\E\Big[F\Big( X^g_\eps+\frac Q2\ln g+\bm c)\Big)e^{-\frac{1}{4 \pi} \int_{\mathbb{R}^2} R_g(x)  \langle Q, X^g_\eps(x)+\bm c \rangle {\rm v}_ g(dx)  -\sum_{i=1}^{r} \mu_i e^{\gamma  \langle e_i, \bm c\rangle }M^{ \gamma e_i,g}_\eps(\C) }\Big]\,d\bm c.
\end{multline*}
The mapping $F\mapsto  \langle F\rangle_{T,g}$ generates a measure on $\mathrm H^{-1}(\R^2\to \green{\mathfrak{a}},g)$. It is worth noticing that this measure has infinite mass because of the $\bm c \rightarrow -\infty$ behaviour of the integrand, but is non trivial as we will see below.

Indeed, there is a special family of functions $F$ that deserve special attention: these are the Vertex Operators which allow to define the correlations function of Toda CFTs. Fix an integer $N\geq 1$ and $N$ distinct points $z_1,\dots,z_N\in \C$ with respective associated weights $\alpha_1,\dots,\alpha_N\in \green{\mathfrak{a}}$. The regularized correlation function is defined by the expression :
\begin{multline}\label{correl_approx}
  \langle V_{\alpha_1,\eps}(z_1) \cdots V_{\alpha_N,\eps}(z_N) \rangle_{T,g}  \coloneqq  \\
  Z(g) \int_{ \green{\mathfrak{a}}}    \E \left [\Big( \prod_{k=1}^N V^g_{\alpha_k,\eps}(z_k)  \Big) e^{-\frac{1}{4 \pi} \int_{\mathbb{R}^2} R_g(x)  \langle Q, X_\eps^g(x)+\bm c \rangle {\rm v}_ g(dx)   - \sum_{i=1}^{r}  \mu_i e^{\gamma  \langle \bm c,e_i\rangle }M_\eps^{ \gamma e_i,g}(\C)}  \right  ]\,d\bm c  .
\end{multline}
The correlation function is then set to be the limit when $\eps\to0$ of the above regularization:
 \begin{equation}\label{correl_limit}
  \langle V_{\alpha_1}(z_1) \cdots V_{\alpha_N}(z_N) \rangle_{T,g} \coloneqq  \underset{\eps \to 0}{\lim}  \langle V_{\alpha_1,\eps}(z_1) \cdots V_{\alpha_N,\eps}(z_N) \rangle_{T,g}.
\end{equation}
The next subsection is devoted to the study of the convergence of this correlation function.

\subsection{Existence of the correlation function}
The form of the correlation function \eqref{correl_approx} is not really convenient when it comes to investigating its convergence as $\eps\to0$. To obtain a reformulation of the correlation functions, we introduce the random measures
\begin{equation}\label{GMC_insert}
 Z_{(\bm z,\bm \alpha)}^{\gamma e_i}  (dx)\coloneqq  e^{\gamma \sum_{j=1}^N  \langle\alpha_j,e_i \rangle G_{\hat{g}} (x,z_j) }  M^{\gamma e_i,\hat g}(dx)  
 \end{equation}
and we define 
\begin{equation}
    \bm s\coloneqq \frac{\sum_{j=1}^N \alpha_j-2Q}\gamma\quad\text{as well as, for all $i$,}\quad
s_i\coloneqq   \frac{\langle \sum_{j=1}^N \alpha_j-2Q,\omega_i^\vee \rangle}{\gamma} \cdot
\end{equation}
The expression of the Vertex Operator as a Wick exponential \eqref{VO_wick} allows us to interpret the product in the regularized correlation function as a Girsanov transform (see Theorem \ref{Girsanov}), and has the effect of shifting the law of the GFF $X^g$. This reformulation is essential to prove the following result:
\begin{theorem}\label{propcorrel1}
Existence and non triviality of the correlation function  $   \langle V_{\alpha_1}(z_1) \cdots V_{\alpha_N}(z_N) \rangle_{T,g} $ do not depend on the background metric $g$ in the conformal class of the spherical metric. Furthermore:
\begin{description}
\item[1. (Seiberg bounds)]  The correlation function $  \langle V_{\alpha_1}(z_1) \cdots V_{\alpha_N}(z_N) \rangle_{T,g} $ exists and is non trivial if and only if the two following conditions hold for all $i=1,\dots,r$: 
\begin{itemize}
\item
$s_i \: >0$,
\item
for all $1\leq k\leq N$, $\langle\alpha_k-Q, e_i \rangle \:  <  \: 0$.
\end{itemize}
\item[2. (Conformal covariance)]  Let $\psi$ be a M\"obius transform of the plane. Then
 \begin{equation*}
 \langle V_{\alpha_1}(\psi (z_1) ) \cdots V_{\alpha_N}(\psi (z_N))   \rangle_{T,g}= \prod_{k=1}^N\norm{\psi'(z_k)}^{-2\Delta_{\alpha_k}}\langle V_{\alpha_1}(z_1) \cdots V_{\alpha_N}(z_N)   \rangle_{T,g}.
 \end{equation*}
where the conformal weights are given by $\Delta_{\alpha_j}\coloneqq \ps{ \frac{\alpha_j}2, Q-\frac{\alpha_j}{2}}$.
\item[3. (Weyl anomaly)]  If $\varphi\in\bar{C}^1(\R^2)$  then we have the following relation
 \begin{equation*}
  \langle V_{\alpha_1}(z_1) \cdots V_{\alpha_N}(z_N)   \rangle_{T,e^\varphi \hat g} =e^{  \frac{\mathbf{c}_T}{96 \pi}S_L(\varphi,\hat g)  }  \langle V_{\alpha_1}(z_1) \cdots V_{\alpha_N}(z_N)   \rangle_{T,\hat{g}}
 \end{equation*}
 where $S_L$ is the Liouville functional 
 $$S_L(\varphi,\hat g)\coloneqq \int_{\R^2 }\big(  |\partial_{\hat{g}} \varphi|^2_{\hat{g}}  + 2   R_{\hat{g}}   \varphi \big)  \,d{\rm v}_{\hat{g}}  ,$$ and the central charge $\mathbf{c}_T$ is given by $\mathbf{c}_T= r + 6 |Q|^2 $.
\item[4. (GMC representation)]  In the particular case where $g=\hat{g}$   is the round metric, one gets the following explicit expression for the correlation function
\begin{multline}\label{expressioncorrels}
 \langle V_{\alpha_1}(z_1) \cdots V_{\alpha_N}(z_N) \rangle_{T,\hat g}     \\
  = \left ( \prod_{i=1}^{r} \frac{\Gamma (s_i)\mu_i^{- s_i}}\gamma   \right )\prod_{k=1}^N \hat g (z_k)^{\Delta_{\alpha_k}}  e^{ \sum_{ k < l} \langle\alpha_k ,\alpha_l \rangle G_{\hat g}(z_k,z_l)  } \E \left [      \prod_{i=1}^{r}   Z_{(\bm z,\bm \alpha)}^{\gamma e_i}  (\C)^{- s_i}   \right ].
\end{multline}
\end{description}
\end{theorem}
 The items 2 and 3 above  characterize the theory as a CFT with central charge $\mathbf{c}_T$, where $\bm c_T$ is given for $\mathfrak g$ simple by:
 \begin{multline}\label{central_charge}
 \arraycolsep=1.9pt\def\arraystretch{1.5}
     \begin{array}{c|c}
        \mathfrak{g} & \bm c_{T,\mathfrak g} \\
        \hline A_{n}& n+\frac{n(n+1)(n+2)}2(\gamma+\frac2\gamma)^2 \\
          B_n & n+ \gamma^2\frac{n(2n-1)(2n+1)}{2}+2n(n+1)(4n-1)\\
          &+\frac{4}{\gamma^2}n(n+1)(2n+1)\\
          C_n& n+\gamma^2\frac{n(n+1)(2n+1)}{2}+2n(n+1)(4n-1)\\
          &+\frac{4}{\gamma^2}{n(2n-1)(2n+1)}\\
          D_n& n+ (n-1)n(2n-1)(\gamma+\frac2\gamma)^2\\
     \end{array}
     \quad
     \arraycolsep=1.9pt\def\arraystretch{1.5}
     \begin{array}{c|c}
        \mathfrak{g} & \bm c_{T,\mathfrak g} \\
        \hline E_6 & 6+468(\gamma+\frac2\gamma)^2\\
          E_7 & 7+1197 (\gamma+\frac2\gamma)^2\\
          E_8 & 8+3720 (\gamma+\frac2\gamma)^2\\
          F_4 & 4+ 234\gamma^2+330+468\frac{4}{\gamma^2}\\
          G_2 & 2+28\gamma^2+192+84\frac{4}{\gamma^2}\\
          &
     \end{array}
 \end{multline}
\begin{proof}
The proof follows closely that developed in~\cite{DKRV} for the case where $\mathfrak{g}=\mathfrak{sl}_2$, that is when the CFT being studied is Liouville CFT. The main difference lies in the fact that many of the quantities involved are no longer scalar but rather vectors of the Euclidean space $\mathfrak{a}$. Likewise several GMC measures, stemming from the form of the Toda action~\eqref{action}, need to be considered.

We start with the first and fourth items. To prove these items we rely on Lemma~\ref{lemmaSeibergplus} from Section~\ref{section:moment} below, which provides sufficient conditions ensuring that the expression~\eqref{expressioncorrels} does make sense probabilistically speaking. Anticipating on the conformal anomaly formula, we can assume that we work with the spherical metric $\hat g$, which is such that \green{
\[{\frac{1}{4 \pi} \int_{\mathbb{C}} \langle Q+X^{\hat g}, \bm c \rangle R_{\hat g}  (x) v_{\hat g}(dx)=2\ps{Q,\bm c}}\] since $X^{\hat g}$ has zero mean value in the metric $\hat g$. Under such an assumption the expressions} of the Vertex operators $V_{\alpha_k,\eps}(z_k)$ as Wick exponentials \eqref{VO_wick} allow to interpret them as Girsanov weights that have the effect of shifting the law of the GFF by an additive term. More precisely, it follows from Theorem~\ref{Girsanov} that 
\begin{align}\label{eq:girsanov_VO}
\expect{\prod_{k=1}^N V^g_{\alpha_k,\eps}(z_k) F(X_\eps^{\hat g})}=\expect{F\left(X_\eps^{\hat g}+\sum_{k=1}^N\alpha_kG_{\hat g,\eps}(\cdot,z_k)\right)}
\end{align}
where $G_{\hat{g},\eps}$, the covariance kernel of $X^{\hat g}_\eps$, is a mollified version of $G_{\hat{g}}$.
This allows to rewrite the regularized correlation function as 
\[
\prod_{k=1}^N \hat g (z_k)^{\Delta_{\alpha_k}}e^{ \sum_{ k < l} \langle\alpha_k ,\alpha_l \rangle G_{\hat g,\eps}(z_k,z_l)  }\int_{ \R^{r}}    e^{\sum_{i=1}^{r} \green\gamma s_i c_i}\E \left [ e^{- \sum_{i=1}^{r} \mu_i e^{\gamma \green{ c_i} }Z_{(\bm z,\bm \alpha),\eps}^{\gamma e_i}  (\C)}  \right  ]\,dc_1...dc_{r}
\] where $Z_{(\bm z,\bm \alpha),\eps}^{\gamma e_i}  (dx)\coloneqq e^{\gamma \sum_{k=1}^N  \langle\alpha_k,e_i \rangle G_{\hat{g},\eps} (x,z_j) }  M^{\gamma e_i,\hat g}(dx)  $. Now, if one of the $s_i$ is non-positive, then the whole integral can be lower-bounded by 
\[
\int_{ \R^{r}}    e^{\sum_{i=1}^{r}\green\gamma s_i c_i} e^{-\sum_{i=1}^{r} \mu_i e^{\gamma  \green{c_i} }M}\,dc_1...dc_{r}\mathbb{P}\left(\forall 1\leq i\leq r,Z_{(\bm z,\bm \alpha),\eps}^{\gamma e_i}  (\C) \leq M\right)=+\infty
\] where $M>0$ is taken so that $\mathbb{P}\left(\forall 1\leq i\leq r,Z_{(\bm z,\bm \alpha),\eps}^{\gamma e_i}  (\C) \leq M\right)>0$ (to see why this is possible note that $G_{\hat{g},\eps}(x,\cdot)$ is bounded over $\C$ and apply Lemma \ref{lemmaSeibergplus} with all the $\alpha$ taken equal to zero).
Therefore the $\eps$-regularized partition function is infinite if one of the $s_i$ is non-positive. Conversely if these $s_i$ are all positive and using Lemma \ref{lemmaSeibergplus} we can make the change of variable $y_i=\mu_i e^{\gamma c_i}Z_{(\bm z,\bm \alpha),\eps}^{\gamma e_i}  (\C)$ \green{for $1\leq i\leq r$} in the integral so that we are left with
\begin{align*}
\prod_{k=1}^N \hat g (z_k)^{\Delta_{\alpha_k}}e^{ \sum_{ k < l} \langle\alpha_k ,\alpha_l \rangle G_{\hat g,\eps}(z_k,z_l)  }\int_{ (0,\infty)^{r}}    \E \left [ \prod_{i=1}^{r} \frac{\left(\mu_i Z_{(\bm z,\bm \alpha),\eps}^{\gamma e_i}  (\C)\right)^{- s_i}}\gamma y_i^{s_i-1}e^{- y_i}  \right  ]\,dy_1...dy_{r}.
\end{align*}
Using Fubini-Tonelli's theorem the latter can be evaluated and is found to be equal to
\begin{align*}
&\prod_{k=1}^N \hat g (z_k)^{\Delta_{\alpha_k}}e^{ \sum_{ k < l} \langle\alpha_k ,\alpha_l \rangle G_{\hat g,\eps}(z_k,z_l)  }\E \left [      \prod_{i=1}^{r} \frac{\left(\mu_i Z_{(\bm z,\bm \alpha),\eps}^{\gamma e_i}  (\C)\right)^{- s_i}}\gamma     \right ]\int_{ (0,\infty)^{r}}  \prod_{i=1}^{r} y_i^{s_i-1}e^{- y_i}  \,dy_1...dy_{r}\\
&=\left ( \prod_{i=1}^{r} \frac{\Gamma (s_i)\mu_i^{- s_i}}\gamma   \right )\prod_{k=1}^N \hat g (z_k)^{\Delta_{\alpha_k}}  e^{ \sum_{ k < l} \langle\alpha_k ,\alpha_l \rangle G_{\hat g,\eps}(z_k,z_l)  } \E \left [      \prod_{i=1}^{r}   Z_{(\bm z,\bm \alpha),\eps}^{\gamma e_i}  (\C)^{- s_i}   \right ].
\end{align*}
To conclude for the first and fourth item it remains to show that: \begin{itemize}
    \item If for all $1\leq i \leq r$, and $1\leq j\leq N$, $\ps{\alpha_j,e_i}<\ps{Q,e_i}$, then 
\begin{equation*}
    \lim\limits_{\eps\rightarrow0}\E \left [      \prod_{i=1}^{r}   Z_{(\bm z,\bm \alpha), \eps}^{\gamma e_i}  (\C)^{- s_i}   \right ]=\E \left [      \prod_{i=1}^{r}   Z_{(\bm z,\bm \alpha)}^{\gamma e_i}  (\C)^{- s_i}   \right ]>0.
\end{equation*}
 \item If for some $1\leq i \leq r$ and $1\leq k\leq N$, $\ps{\alpha_k,e_i}\geq\ps{Q,e_i}$, then 
\begin{equation*}
    \lim\limits_{\eps\rightarrow0}\E \left [      \prod_{i=1}^{r}   Z_{(\bm z,\bm \alpha), \eps}^{\gamma e_i}  (\C)^{- s_i}   \right ]=0.
\end{equation*}
\end{itemize}
Let us assume that for all $1\leq i \leq r$, and $1\leq k\leq N$, $\ps{\alpha_k,e_i}<\ps{Q,e_i}$. Then we know from Lemma \ref{lemmaSeibergplus} that the family of random variables $\left(\prod_{i=1}^{r}   Z_{(\bm z,\bm \alpha), \eps}^{\gamma e_i}  (\C)^{- s_i}\right)_{\eps\geq0}$ have (uniformly bounded in $\eps$) positive moments of all orders. Thus we can write that
\begin{align*}
    &\E \left [ \norm{     \prod_{i=1}^{r}   Z_{(\bm z,\bm \alpha), \eps}^{\gamma e_i}  (\C)^{- s_i}  - \prod_{i=1}^{r}   Z_{(\bm z,\bm \alpha)}^{\gamma e_i}  (\C)^{- s_i} }  \right ]\leq \E \left [ \norm{   Z_{(\bm z,\bm \alpha), \eps}^{\gamma e_1}  (\C)^{- s_1}-Z_{(\bm z,\bm \alpha)}^{\gamma e_1}  (\C)^{- s_1}}  \prod_{i=2}^{r}   Z_{(\bm z,\bm \alpha), \eps}^{\gamma e_i}  (\C)^{- s_i}   \right ]\\
    & +\E \left [\norm{     \prod_{i=2}^{r}   Z_{(\bm z,\bm \alpha), \eps}^{\gamma e_i}  (\C)^{- s_i}  - \prod_{i=2}^{r}   Z_{(\bm z,\bm \alpha)}^{\gamma e_i}  (\C)^{- s_i} } Z_{(\bm z,\bm \alpha)}^{\gamma e_1}  (\C)^{- s_1}  \right ]\\
    &\leq \E \left [ \norm{   Z_{(\bm z,\bm \alpha), \eps}^{\gamma e_1}  (\C)^{- s_1}-Z_{(\bm z,\bm \alpha)}^{\gamma e_1}  (\C)^{- s_1}}^p\right]^{\frac1p}\E\left[\prod_{i=2}^{r}   Z_{(\bm z,\bm \alpha), \eps}^{\gamma e_i}  (\C)^{- qs_i}   \right ]^{\frac 1q}\\
    &+\E \left [\norm{     \prod_{i=2}^{r}   Z_{(\bm z,\bm \alpha), \eps}^{\gamma e_i}  (\C)^{- s_i}  - \prod_{i=2}^{r}   Z_{(\bm z,\bm \alpha)}^{\gamma e_i}  (\C)^{- s_i} }^p\right]^{\frac1p}\E\left[ Z_{(\bm z,\bm \alpha)}^{\gamma e_1}  (\C)^{- qs_1}  \right ]^{\frac1q}
\end{align*}
where we have used H\"older inequality with some $p=\frac{q}{q-1}>1$. Therefore we can proceed by induction on $r$ so that the only point to check is ${\lim\limits_{\eps\to0}\E \left [ \norm{   Z_{(\bm z,\bm \alpha), \eps}^{\gamma e_1}  (\C)^{- s_1}-Z_{(\bm z,\bm \alpha)}^{\gamma e_1}  (\C)^{- s_1}}^p\right]^{\frac1p}=0}$. This fact has already been proved by the authors in \cite[Lemma 3.3]{DKRV}.
For the second bullet point, let us introduce the set ${\mathcal{P}\coloneqq \{i=1,\dots ,r\,\vert\exists1\leq k\leq N,\green{\ps{\alpha_k-Q,e_i}\geq 0}\}}$ and assume that it is non-empty. Then we can write that, for positive $\frac 1{p_i}$ and $\frac 1q$ summing to one,
\[
    \E \left [      \prod_{i=1}^{r}   Z_{(\bm z,\bm \alpha), \eps}^{\gamma e_i}  (\C)^{- s_i}   \right ]\leq \prod_{i\in\mathcal{P}}\E \left [      Z_{(\bm z,\bm \alpha), \eps}^{\gamma e_i}  (\C)^{- p_is_i}   \right ]^{\frac1{p_i}}\E \left [      \prod_{i\not\in\mathcal{P}}   Z_{(\bm z,\bm \alpha), \eps}^{\gamma e_i}  (\C)^{- qs_i}   \right ]^{\frac1q}.
\]
Then we have already seen that the second expectation in the right-hand-side had a finite limit as $\eps\rightarrow0$ thanks to the results of Lemma \ref{lemmaSeibergplus}. Conversely standard results of the GMC theory (see again \cite[Lemma 3.3]{DKRV} for instance) imply that for any $i\in\mathcal{P}$, ${\lim\limits_{\eps\to0}\E \left [     Z_{(\bm z,\bm \alpha), \eps}^{\gamma e_i}  (\C)^{- p_is_i}   \right ]^{\frac1{p_i}}=0}$. This proves the first item in the statement of Theorem~\ref{propcorrel1}.

We now turn to the proofs of the second and third items.
Let us start with the third item. Since in the definition of the correlation function~\eqref{correl_approx} the expressions involved actually depend on $X^g +\bm c$ rather than $X^g$, by making the change of variable in the zero mode $\bm c$ given by $\bm c\leftrightarrow \bm c-m_{\hat g}(X^g)$ we can in fact rewrite Equation~\eqref{correl_approx} by replacing the field $X^g+\bm c$ by $X^{g}-m_{\hat g}(X^g) +\bm c$. Using the fact that $X^g-m_{\hat g}(X^g)$ and $X^{\hat g}$ have same law, we can therefore assume that $X$ has the law of a GFF with vanishing mean with respect to the round metric $X^{\hat g}$. Since $g=e^{\varphi} \hat{g}$, we have that
 \begin{multline*}
 \langle  V_{\alpha_1,\eps}(z_1) \cdots V_{\alpha_N,\eps}(z_N)   \rangle_g =  Z(g)\prod_{k=1}^N e^{\frac{\ps{\alpha_k,Q}}{2}\varphi(x_k)}  \times \underset{\eps \to 0}{\lim}  \\
  \int_{\green{\mathfrak a}}     e^{\gamma\ps{\bm s,\bm c}}     \E \left [ \prod_{j=1}^k \tilde V^{\hat g}_{\alpha_j,\eps}(z_j)   e^{-\frac{1}{4 \pi} \int_{\mathbb{C}} R_g  \langle Q, X \rangle (x) v_g(dx)  - \sum_{i=1}^{r} \mu_i e^{\gamma c_i} \int_{\mathbb{C}}  e^{  \gamma^2 \frac{\varphi(x)}{2}  }  \green{V^{\hat g}_{\gamma e_i,\eps}}(x)  v_g(dx) }  \right  ]  d\bm c
 \end{multline*}  
 where this time regularization is done with respect to the round metric, and $\tilde V$ is the \green{Vertex Operator} without constant mode, that is 
 \[
 \tilde V^g_{\alpha,\eps}(z)\coloneqq  \eps^{ \frac{|\alpha|^2}{2}}  e^{  \langle\alpha, X^g_\eps(z) + \frac Q2\ln g-\frac{\alpha\theta_{\eta}}2\rangle}.
\]
We will consider the term $e^{-\frac{1}{4 \pi} \int_{\mathbb{C}} R_g  \langle Q, X \rangle (x) v_g(dx)}$ as a Girsanov transform. Namely we can use the fact that $R_g(y)v_g(dy)=(-\Delta_{\hat{g}} \varphi(y) +2)v_{\hat g}(dy)$ (at least in the weak sense since $\varphi\in \bar{C}^1(\R^2)$) and the definition of the Green function $G_{\hat g}$ to see that for any $\alpha\in\mathfrak{a}$: 
\begin{equation*}
 \E \left [  \left (  \frac{1}{4\pi}\int_{\mathbb{C}} R_g(y)  \langle Q, X \rangle (y) v_g(dy)  \right )   \langle \alpha, X\rangle (x) \right ]=   \langle Q,\frac\alpha2 \rangle  \left( \varphi(x)  -m_{\hat g}(\varphi)\right),
\end{equation*}
The variance of this expression is given by
 \begin{align*}
 & \E \left [  \left ( \frac{1}{4\pi} \int_{\C} R_g(x)  \langle Q, X \rangle (x)  v_g(dx) \right )^2 \right ]    \\
 &  =  \frac{1}{16\pi^2} \int_{\C \times \C} R_g(x) R_g(y)  \E[  \langle Q, X \rangle (x)  \langle Q, X \rangle (y)   ]  v_g(dx)v_g(dy)   \\
& = \frac{1}{16\pi^2}|Q|^2 \int_{\C \times \C} R_g(x) R_g(y)  G_{\hat g}(x,y) v_g(dx)v_g(dy)    \\ 
& = \frac{1}{16\pi^2} |Q|^2 \int_{\C } R_g(x)  \left ( \int_{\C}  R_g(y)  G_{\hat g}(x,y)v_g(dy)   \right ) v_g(dx)         \\
& = \frac{1}{8\pi}|Q|^2 \int_{\C } R_g(x) \left(\varphi(x)  -m_{\hat g}(\varphi)\right)  v_g(dx) \\
& = \frac{1}{8\pi}|Q|^2 \int_{\C } \left(-\Delta_{\hat g} \varphi(x)+2\right)  \left(\varphi(x)  -m_{\hat g}(\varphi)\right)  v_{\hat g}(dx) \\
& =  \frac{1}{8\pi} |Q|^2 \int_{\C }  |\partial_{\hat{g}} \varphi|^2  v_{\hat{g}}(dx). 
 \end{align*}
As a consequence we obtain thanks to Theorem~\ref{Girsanov} that 
\[ 
	\expect{e^{-\frac{1}{4 \pi} \int_{\mathbb{C}} R_g  \langle Q, X \rangle (x) v_g(dx)}F(X)}=e^{\frac{1}{16\pi} |Q|^2 \int_{\C }  |\partial_{\hat{g}} \varphi|^2  v_{\hat{g}}(dx)}\expect{F\left(X+\frac{Q}{2}\left( \varphi(x)  -m_{\hat g}(\varphi)\right)\right)}.
\]
Using like before the change of variable $\bm c\leftrightarrow\bm c +\frac Q2m_{\hat g}(\varphi)$ and recollecting terms we end up with
\begin{align*}
  \displaystyle\langle V_{\alpha_1}(z_1) &\cdots V_{\alpha_N}(z_N)   \rangle _g= Z(g)e^{  \frac{|Q|^2}{16 \pi} \int_{\C }  |\partial_{\hat{g}} \varphi|^2 v_{\hat{g}} } \times\\
  &\underset{\eps \to 0}{\lim}\int_{\green{\mathfrak a}}    e^{\gamma\ps{\bm s,\bm c}}    \E \left [ \prod_{k=1}^N \tilde V^{\hat g}_{\alpha_k,\eps}(z_k)   e^{ - \sum_{i=1}^{r} \mu_i e^{\gamma c_i} \int_{\C}  e^{ \varphi(x) (\gamma^2 \frac{|e_i|^2}{4}-\frac{\gamma  \langle Q,e_i \rangle }{2})  }  \green{\tilde V_{\gamma e_i,\eps}}(x)   v_g(dx) }  \right  ]  d\bm c.
\end{align*}
Since thanks to Equation~\eqref{eq:propQ} we know that for all $1\leq i\leq r$, $\frac{|\gamma e_i|^2}{4}-\frac{\gamma  \langle Q,e_i \rangle }{2}=-1$, by the change of variable ${\green{\bm c \leftrightarrow \bm c-  \frac{Q}2 m_{\hat g}(\varphi)}}$ we get that
  \begin{align*}
 &\langle V_{\alpha_1}(z_1) \cdots V_{\alpha_N}(z_N)   \rangle _g =  \\
 &Z(g) e^{  \frac{|Q|^2}{16 \pi} \int_{\C }  |\partial_{\hat{g}} \varphi|^2  v_{\hat{g}} + \frac{|Q|^2}{4 \pi}   \int_{\C }  \varphi v_{\hat{g}} } \times  \underset{\eps \to 0}{\lim}\int_{\green{\mathfrak{a}}} \E \left [ \prod_{k=1}^N V^{\hat g}_{\eps, \alpha_k}(z_k)   e^{ -\sum_{i=1}^r \mu_i \int_{\C}   \green{V^{\hat g}_{\gamma e_i,\eps}}(x)   v_{\hat{g}}(dx) }  \right  ]d\bm c,  
 \end{align*}   
whence the result, by using the expression \eqref{reg_det} for the regularized determinant $Z(g)$.
 
For the second item, we see that according to our proof of the first item \green{and more precisely Equation~\eqref{eq:girsanov_VO}, the quantity  $\ps{F\prod_{k=1}^NV_{\alpha_k}(z_k)}$ for $F$ bounded continuous over $\mathrm{H}^{-1}(\S^2\to\green{\mathfrak{a}},\hat g)$ is actually given by
\begin{align*}
Z(g)\prod_{k=1}^N \hat g &(z_k)^{\Delta_{\alpha_k}}e^{ \sum_{ k < l} \langle\alpha_k ,\alpha_l \rangle G_{\hat g}(z_k,z_l)  }\\
&\int_{ \green{\mathfrak{a}}}    e^{\gamma \ps{\bm s,\bm c}}\E \left [ F\left(\green{X^{\hat g}}+\frac Q2\ln\hat g+\bm c+\sum_{k=1}^N\alpha_kG_{\hat g}(\cdot,z_k)\right)e^{- \sum_{i=1}^{r} \mu_i e^{\gamma  \langle \bm c,e_i\rangle }Z_{(\bm z,\bm \alpha)}^{\gamma e_i}  (\C)}  \right  ]\,d\bm c.
\end{align*}
Because} $\psi$ is a conformal map we know that the Riemannian metric $\psi^*\hat g$ (that we have denoted $\hat{g}_\psi$ before) \green{lies within} the conformal class of $\hat g$; as a consequence the \green{GFFs} $\green{X^{\hat g_\psi}}-m_{\hat g}(\green{X^{\hat g_\psi}})$ and $\green{X^{\hat g}}$ have same law. Moreover from Lemma \ref{conformal_covariance_green} we know that $\green{X^{\hat g_\psi}}$ has same law as $\green{X^{\hat g}}\circ\psi$. In a nutshell, 
\[
X^{\hat g}\circ\psi-m_{\hat g}(X^{\hat g}\circ\psi)\eqlaw X^{\hat g}.
\]
Besides we saw in Equation~\eqref{conformal_spherical} that $G_{\hat g}(\cdot,\psi(z_k))\circ\psi+\frac14(\phi+\phi(z_k))=G_{\hat g}(\cdot,z_k)$ where $e^{\phi}=\frac{\hat g_\psi}{\hat g}$. Combining these two assertions yields that the laws of 
\[\green{X^{\hat g}}+\sum_{k=1}^N\alpha_kG_{\hat g}(\cdot,z_k)\quad\text{and }\quad\left(\green{X^{\hat g}}+\sum_{k=1}^N\alpha_kG_{\hat g}(\cdot,\green{\psi(z_k)})\right)\circ\psi+\frac14\sum_{k=1}^N\alpha_k(\phi+\phi(z_k))-m_{\hat g}(\green{X^{\hat g}}\circ\psi)
\]
are actually the same.
Since $\hat g_\psi=\norm{\psi'}^2\hat g\circ\psi$ the latter further implies that 
\[
\bm\Phi_{\hat g}^{\bm z}\eqlaw \bm\Phi_{\hat g}^{\psi\bm z}\circ\psi+ Q\ln\norm{\psi'}-\frac Q2\phi+\frac14\sum_{k=1}^N\alpha_k(\phi+\phi(z_k))-m_{\hat g}(X^{\hat g}\circ\psi)
\]
where $\bm\Phi_{\hat g}^{\bm z}$ is a shorthand for $X^{\hat g}+\frac Q2\ln\hat g+\sum_{k=1}^N\alpha_kG_{\hat g}(\cdot,z_k)$ and $\psi\bm z\coloneqq (\psi(z_1),\cdots,\psi(z_N))$. Therefore $\ps{F\prod_{k=1}^NV_{\alpha_k}(z_k)}$ can be put under the form
\begin{align*}
&Z(g)\prod_{k=1}^N \hat g (z_k)^{\Delta_{\alpha_k}}e^{ \sum_{ k < l} \langle\alpha_k ,\alpha_l \rangle G_{\hat g}(z_k,z_l)  }\int_{ \mathfrak{a}}    e^{\gamma \ps{\bm s,\bm c}}\,d\bm c\\
&\E \left [ F\left(\bm\Phi_{\hat g}^{\psi\bm z}\circ\psi+ Q\ln\norm{\psi'}+\frac{\gamma \bm s}{4}\phi+\bm c+\frac14\sum_{k=1}^N\alpha_k\phi(z_k)-m_{\hat g}(X^{\hat g}\circ\psi)\right)e^{- \sum_{i=1}^{r} \mu_i e^{\gamma  \langle \bm c,e_i\rangle }Z_{(\bm z,\bm \alpha)}^{\gamma e_i}  (\C)}  \right ].
\end{align*}
This motivates the shift in the zero mode $\bm c\leftrightarrow\bm c+ m_{\hat g}(X^{\hat g})-\frac14\sum_{k=1}^N\alpha_k\phi(z_k)-\frac\gamma4\bm sm_{\hat g}(\phi)$. After this change of variable we are left with
\begin{align*}
&Z(g)\prod_{k=1}^N \hat g (z_k)^{\Delta_{\alpha_k}}e^{ \sum_{ k < l} \langle\alpha_k ,\alpha_l \rangle G_{\hat g}(z_k,z_l)  }e^{-\frac{\gamma}{4}\ps{\bm s,\alpha_k}}\int_{ \mathfrak{a}}    e^{\gamma \ps{\bm s,\bm c}}\,d\bm c \\
&\E \left [ e^{\gamma\ps{\bm s,m_{\hat g}(X^{\hat g})}-m_{\hat g}(\phi)\frac{\gamma^2\ps{\bm s,\bm s}}4}F\left(\bm\Phi_{\hat g}\circ\psi+ Q\ln\norm{\psi'}+\frac14\bm s(\phi-m_{\hat g}(\phi))\right)e^{- \sum_{i=1}^{r} \mu_i e^{\gamma  \langle \bm c,e_i\rangle }Z_{(\bm z,\bm \alpha)}^{\gamma e_i}  (\C)}  \right  ].
\end{align*}
Collecting up terms using \eqref{conformal_spherical} yields:
\begin{align*}
&\green{Z(g)}\prod_{k=1}^N \hat g_\psi (z_k)^{\Delta_{\alpha_k}}e^{ \sum_{ k < l} \langle\alpha_k ,\alpha_l \rangle G_{\hat g}(\psi(z_k),\psi(z_l))  }\int_{ \green{\mathfrak{a}}}    e^{\green\gamma\ps{\bm s,\bm c}}\,d\bm c\\
&\E \left [e^{\green\gamma\ps{\bm s,m_{\hat g}(\green{X^{\hat g}})}-m_{\hat g}(\phi)\frac{\gamma^2\ps{\bm s,\bm s}}4} F\left(\bm\Phi_{\hat g}\circ\psi+ Q\ln\norm{\psi'}+\frac14\bm s(\phi-m_{\hat g}(\phi))\right)e^{- \sum_{i=1}^{r} \mu_i e^{\gamma  \langle \bm c,e_i\rangle }Z_{(\bm z,\bm \alpha)}^{\gamma e_i}  (\C)}  \right  ].
\end{align*}
\green{The proof is completed by interpreting the exponential term $e^{\ps{\bm s,m_{\hat g}(\green{X^{\psi^*\hat g}})}-\frac{m_{\hat g}(\phi)\ps{\bm s,\bm s}}4}$ as a Girsanov transform, whose effect is to} shift the law of $\bm\Phi_{\hat g}\circ\psi$ by $-\frac14\bm s(\phi-m_{\hat g}(\phi))$.
\end{proof}
In the proof of the conformal covariance property we have shown a slightly more general result. Indeed we have proved that under the Seiberg bounds and for any M\"obius transform of the plane, the following was true for any continuous and bounded map $F$ on $\mathrm{H}^{-1}(\S^2\to\green{\mathfrak{a}},\hat g)$:
\begin{equation}
    \ps{F\left(\green{\cdot\circ\psi}+Q\ln\norm{\psi'}\right)\prod_{k=1}^NV_{\alpha_k}(\psi(z_k))}_g=\prod_{k=1}^N\norm{\psi'(z_k)}^{-2\Delta_{\alpha_k}}\ps{\green F\prod_{k=1}^NV_{\alpha_k}(z_k)}_{g}.
\end{equation}
This statement is usually referred to as the conformal covariance of the Toda field.

 %%%%%%%%%%%%%%%%%%%%%%%%%%%%%%%%%%%%%%%%%%%%%%%%%%%%%%%%%%%%%%%%%%%%%%%%%%%%%%%%%%%%%%%%%%%%%%%%%%%%%%%%%%%%%%%%%%%%%%%%%%%%%%%%%%%%%%%%%%%%%%%%%%%%%%%%%%%%%%%%%%%%%%%%%%%%%%%%%%%%%%%%%%%%%%%%%%%%%%
 \section{A moment bound}\label{section:moment}
  %%%%%%%%%%%%%%%%%%%%%%%%%%%%%%%%%%%%%%%%%%%%%%%%%%%%%%%%%%%%%%%%%%%%%%%%%%%%%%%%%%%%%%%%%%%%%%%%%%%%%%%%%%%%%%%%%%%%%%%%%%%%%%%%%%%%%%%%%%%%%%%%%%%%%%%%%%%%%%%%%%%%%%%%%%%%%%%%%%%%%%%%%%%%%%%%%%%%%

We wish to extend here the validity of the probabilistic representation  \eqref{expressioncorrels}. The point is that  the explicit expression \eqref{expressioncorrels} allows us to isolate the constraints $s_i>0$ in the product of $\Gamma$ functions. This term can obviously be analytically removed. The question is then to determine whether the expectation in  \eqref{expressioncorrels} makes sense beyond the range of parameters permitted by the Seiberg bounds. So we claim
   
 \begin{lemma}[Extended Seiberg bounds]\label{lemmaSeibergplus}
The bound 
\begin{equation}\label{expectunit}
\E \left [      \prod_{i=1}^{r}  ( Z_{(\bm z,\bm \alpha)}^{\gamma e_i}  (\R^2))^{- s_i}   \right ]  < \infty
\end{equation}
holds if and only if for all $i=1,\cdots,r$ one has
\begin{equation}\label{extSeiberg}
  -s_i < \frac{4}{\gamma^2\ps{e_i,e_i}}\wedge \min_{k=1,\dots,N} \frac{1}{\gamma}\langle Q-\alpha_k,e_i^\vee\rangle
\end{equation}
\end{lemma} 
 \proof
We suppose that condition  \eqref{extSeiberg} holds. Let us consider the families of indices $$\mathcal{P}\coloneqq \{i=1,\dots ,r\,|\,s_i\geq 0\}\quad \text{ and }\quad \mathcal{N}\coloneqq \{i=1,\dots ,r\,|\,s_i<0\}.$$ 
Choose $p>1$ such that for all $i\in \mathcal{N}$, $-p s_i< \frac{2}{\gamma^2}\wedge \min\limits_{k=1,\dots,N} \frac{1}{\gamma}\langle Q-\alpha_k,e_i^\vee\rangle$ and fix the conjugate exponent $q>1$ such that $\tfrac{1}{p}+\tfrac{1}{q}=1$. By H\"older inequality we can write that
\begin{align*}
\E \left [      \prod_{i=1}^{r}  ( Z_{(\bm z,\bm \alpha)}^{\gamma e_i}  (\R^2))^{- s_i}   \right ]  \leq \E \left [      \prod_{i\in \mathcal{P}}  ( Z_{(\bm z,\bm \alpha)}^{\gamma e_i}  (\R^2))^{-\green{ qs_i}}   \right ]^{\green{1/q}}  \E \left [      \prod_{i\in \mathcal{N}}    ( Z_{(\bm z,\bm \alpha)}^{\gamma e_i}  (\R^2))^{- \green{p s_i}}   \right ]^{\green{1/p}}.
\end{align*} 
The product running over $i\in \mathcal{P}$ is finite because GMC admits negative moments of all order (see \cite[Theorem 2.12]{review}). For the  product running over $i\in \mathcal{N}$, we use Corollary \ref{comparison} in appendix as well as the relation \eqref{covX}, which shows that the GFFs $\langle \gamma e_i,\green{X^g}\rangle$ and $\langle \gamma e_j,\green{X^g}\rangle$, for $i\not=j$, are negatively correlated since $\langle \gamma e_i,\gamma e_j\rangle=\gamma
^2A_{i,j}$ (recall that $A$ is the Cartan matrix) and all off-diagonal elements of $A$ are nonpositive. Hence
\begin{equation}\label{ineg1}
   \E \left [      \prod_{i\in \mathcal{N}}    ( Z_{(\bm z,\bm \alpha)}^{\gamma e_i}  (\R^2))^{- \green{p s_i}}   \right ] \leq      \prod_{i\in \mathcal{N}} \E \left [       ( Z_{(\bm z,\bm \alpha)}^{\gamma e_i}  (\R^2))^{- \green{p s_i}}   \right ] .
\end{equation} 
The GMC measure which appears in the expression of $Z_{(\bm z,\bm \alpha)}^{\gamma e_i}$ is defined from the GFF $\ps{X^g,e_i}$. This GFF has the law of $\norm{e_i}X^g_0$ where we have denoted by $X^g_0$ the real-valued field considered in~\cite[ Lemma A.1]{DKRV}. This amounts to replacing the coupling constant $\gamma$ by $\gamma_i\coloneqq \gamma\norm{e_i}$ and the weight of the insertion by $\alpha_k^i\coloneqq \ps{\alpha_k,\frac{e_i}{\norm{e_i}}}$ in the statement of~\cite[ Lemma A.1]{DKRV}. This entails that the corresponding expectation term is finite provided that $-ps_i < \frac{4}{\gamma_i^2}\wedge \min_{k=1,\dots,N} \frac{2}{\gamma_i}(\frac{\gamma_i}2+\frac2{\gamma_i}-\alpha_k^i)$. The latter can be rewritten under the form \[
-ps_i < \frac{4}{\gamma \ps{e_i,e_i}}\wedge \min_{k=1,\dots,N} \frac{1}{\gamma}\left(\gamma+\frac4{\gamma\ps{e_i,e_i}}-\ps{\alpha_k,\frac{2e_i}{\ps{e_i,e_i}}}\right).
\]
We conclude with the help of Equation~\eqref{eq:propQ} that each expectation in the product above is finite thanks to our assumptions on the $ps_i$, $1\leq i\leq r$.

Conversely, assume that the expectation \eqref{expectunit} is finite. By Corollary \ref{comparison} in the appendix applied to the function 
$$H(x_1,\dots,x_r)=  -\prod_{i\in \mathcal{N}} x_i^{- s_i} \prod_{i\in \mathcal{P}} x_i^{- s_i} , $$ with the partition $(\mathcal{P},\mathcal{N})$ of $\{1,\cdots,r\}$ and to the GFFs $(\langle \gamma e_i,X^g\rangle)_{i=1,\dots,r}$ we deduce
\begin{align*}
\E \left [      \prod_{i=1}^{r}  ( Z_{(\bm z,\bm \alpha)}^{\gamma e_i}  (\R^2))^{- s_i}   \right ]  \geq \E \left [      \prod_{i\in \mathcal{P}}  ( Z_{(\bm z,\bm \alpha)}^{\gamma e_i}  (\R^2))^{-  s_i}   \right ]  \E \left [      \prod_{i\in \mathcal{N}}    ( Z_{(\bm z,\bm \alpha)}^{\gamma e_i}  (\R^2))^{-   s_i}   \right ] .
\end{align*} 
Since GMC admits negative moments of all order \cite[Theorem 2.12]{review}, the first expectation in the right-hand side above is a finite constant $C>0$. This implies that the second expectation is finite too. From now on, we fix $i_0\in \mathcal{N}$ and $j\in\{1,\dots,N\}$. Without loss of generality and for the sake of simplicity, we may assume that $z_j=0$.  Then we can choose $\delta>0$ such that $\min_{j'\not=j}|z_j'|>10\times\delta$   and we can choose non empty balls $(B_i)_{i\not= i_0,i\in \mathcal{N}}$ all of them at distance at least $10\times\delta>0$ from each other and all of them at distance at least $10\times \delta$ from all the $z_j$'s. Set $B_{i_0}\coloneqq B(0,\delta)$. Obviously we have
$$\E \left [      \prod_{i\in \mathcal{N}}    ( Z_{(\bm z,\bm \alpha)}^{\gamma e_i}  (\R^2))^{-   s_i}   \right ] \geq  \E \left [        \prod_{i\in \mathcal{N}}    ( Z_{(\bm z,\bm \alpha)}^{\gamma e_i}  (B_i))^{-   s_i} \right ] .$$
Consider the mean value of the field  $Y\coloneqq \frac{1}{2\pi i}\oint_{\green{|x|=2\delta}}X^g(x)\frac{dx}{x}$. A simple check of covariances shows that the law of the field $X^g-Y$  is the independent sum of the field  $\green{X^g_h}$---which coincides with $\green{X^g}-Y$ outside of $B(0,2\delta)$ and corresponds inside $ B(0,2\delta)$ to the harmonic extension   (component by component) of the field $\green{X^g}-Y$ restricted to  the boundary $ \partial B(0,2\delta)$---plus the Dirichlet field \green{$X_D$} defined by 
\green{\begin{equation}
X_D=\sum_{i=1}^{r}\omega_iX_{D,i},
\end{equation}
where $(X^g_{D,1}, \dots, X^g_{D,r})$} is a family of centered correlated Dirichlet GFFs inside $B(0,2\delta)$ with covariance structure given by
\green{\begin{equation*}
 \E[X^g_{D,i}(z) X^g_{D,j}(z')]=A_{i,j}   G_D(z,z') 
\end{equation*}
and $G_D(z,z')$} stands for the Dirichlet Green function inside $B(0,2\delta)$. From now on we will write $Z_{(\bm z,\bm \alpha)}^{\langle\gamma e_i,X^g\rangle}  (d^2x)$ instead of $Z_{(\bm z,\bm \alpha)}^{\gamma e_i}  (d^2x)$ to indicate in the notations 
the dependence on the underlying Gaussian field.  This means that, generally speaking, we will write $Z_{(\bm z,\bm \alpha)}^{\langle\gamma e_i,X\rangle} $ for
$$Z_{(\bm z,\bm \alpha)}^{\langle\gamma e_i,X\rangle}  (d^2x)\coloneqq \lim_{\eps\to 0}e^{ \sum_{j=1}^N  \langle\alpha_j,\gamma e_i \rangle G_{\hat{g}} (x,z_j) }  \eps^{ \frac{|\gamma e_i|^2}{2}}  e^{  \langle\gamma e_i,  X_\eps(x) \rangle} {\rm v}_g(dx) $$
where $X_\eps$ stands for the $\eps$-regularization of the field $X$ in the metric $g$. So we can write
\begin{align*}
  \E \left [        \prod_{i\in \mathcal{N}}    ( Z_{(\bm z,\bm \alpha)}^{\langle\gamma e_i,X^g\rangle}  (B_i))^{-   s_i} \right ] &=  \E \left [   e^{-\sum_{i\in\mathcal{N}}s_i\langle\gamma e_i,Y\rangle}     \prod_{i\in \mathcal{N}}    ( Z_{(\bm z,\bm \alpha)}^{\langle\gamma e_i,X^g-Y\rangle}  (B_i))^{-   s_i} \right ] .
\end{align*} 
We can remove the factor $ e^{-\sum_{i\in\mathcal{N}}s_i\langle\gamma e_i,Y\rangle}$ by viewing it as a Girsanov transform. Namely, denote by $\sigma^2$ the variance of the centered Gaussian random variable $\sum_{i\in\mathcal{N}}s_i\langle\gamma e_i,Y\rangle$. Then by Girsanov theorem we can write that weighting the law of $X^g$ by 
\[
e^{-\sum_{i\in\mathcal{N}}s_i\langle\gamma e_i,Y\rangle-\frac{\sigma^2}{2}}
\]
amounts to shifting $X^g$ by $-\sum_{j\in\mathcal{N}}s_j\gamma e_j\frac{1}{2\pi i}\oint_{|x|=\delta}G_g(\cdot,x)\frac{dx}{x}$. The values of the variance $\sigma^2$ and of the covariance of $Y$ with $X^g-Y$ are actually irrelevant to conclude. Indeed we only need to know that the variance $\sigma ^2$ is bounded, and that the covariance of $Y$ with $X^g(x)-Y$ is uniformly bounded for $x$ inside $B_i$ and for all $i\in\mathcal{N}$. This is readily seen from their definition. This entails the existence of some positive constant $C>0$ such that
\begin{align*}
\E \left [   e^{-\sum_{i\in\mathcal{N}}s_i\langle\gamma e_i,Y\rangle}     \prod_{i\in \mathcal{N}}    ( Z_{(\bm z,\bm \alpha)}^{\langle\gamma e_i,X^g-Y\rangle}  (B_i))^{-   s_i} \right ] \geq &  C \E \left [        \prod_{i\in \mathcal{N}}    ( Z_{(\bm z,\bm \alpha)}^{\langle\gamma e_i,X^g-Y\rangle}  (B_i))^{-   s_i} \right ].
\end{align*} 
Using the decomposition of the law  of \green{$X^g-Y=X_D+X^g_h$} and independence of \green{$X_D$} and $X^g_h$, we  get   
\begin{align*}
  \E \left [        \prod_{i\in \mathcal{N}}    ( Z_{(\bm z,\bm \alpha)}^{\langle\gamma e_i,X^g-Y\rangle}  (B_i))^{-   s_i} \right ]\geq  \E \left [      ( Z_{(\bm z,\bm \alpha)}^{\langle\gamma e_{i_0},\green{X_D}\rangle}  (B_{i_0}))^{-   s_{i_0}} \right ]  \E \left [ e^{\min_{x\in B_{i_0}}X^g_h(x)}       \prod_{i\not =i_0,i\in \mathcal{N}}    ( Z_{(\bm z,\bm \alpha)}^{\langle\gamma e_i,X^g-Y\rangle}  (B_i))^{-   s_i} \right ].
\end{align*} 
This implies that both expectations in the right-hand side are finite (they are obviously nonzero).
%We claim that Lemma \eqref{kah2} entails that
%\begin{multline}\label{ineq2}
%\E \left [     ( Z_{(x_j,\alpha_j)}^{\gamma e_{i_0}}  (B(x_j,\delta/2)))^{-   s_{i_0}}   \prod_{i\in \mathcal{N},i\not=i_0}         ( Z_{(x_j,\alpha_j)}^{\gamma e_i}  (B_i))^{-   s_i}   \right ] \\
%\geq C_\delta\E \left [     ( Z_{(x_j,\alpha_j)}^{\gamma e_{i_0}}  (B(x_j,\delta/2)))^{-   s_{i_0}} \right] \prod_{i\in \mathcal{N},i\not=i_0}  \E\left[         ( Z_{(x_j,\alpha_j)}^{\gamma e_i}  (B_i))^{-   s_i}   \right ] 
%\end{multline} 
Finiteness of the first expectation above entails, like before by adapting~\cite[ Lemma A.1]{DKRV}, that  $-s_{i_0}<\frac{4}{\gamma^2\ps{e_i,e_i}}\wedge  \frac{1}{\gamma}\langle Q-\alpha_j,e_i^\vee\rangle$. Since the argument is valid for all $i_0\in\mathcal{N}$ and all $j\green{\in\{1,\cdots,N\}}$, this yields the result.
%
%To see why \eqref{ineq2} holds, define
%$$c_\delta\coloneqq \min_{i\not =i'\in \mathcal{N}}\min_{x_i\in B_i,x_{i'}\in B_{i'}}\E\big[\langle \gamma e_i,X_g(x_i)\rangle \langle \gamma e_{i'},X_g(x_{i'})\rangle\big].$$
%Consider a sequence of independent standard Gaussian random variables $(Z_i)_{i\in\mathcal{N}}$ and another Gaussian random variable $Z$, all of them independent from everything. Let $(Y_i)_{i\in\mathcal{N}}$ be a sequence of independent Gaussian fields, with $ Y_i$ having the same law as $\langle \gamma e_i,X_g\rangle$ for all $i\in\mathcal{N}$. Then, for all $i\not=i'\in\mathcal{N}$,  for all $x\in B_i$ and all $x'\in B_{i'}$,
%\begin{multline}
% \E\big[\big(|c_\delta|^{1/2} Z_i+Y_i(x)  \big)\big(|c_\delta|^{1/2} Z_{i'}+Y_{i'}(x') \big)\big]\\
% \leq \E\big[\big(|c_\delta|^{1/2} Z+\langle \gamma e_i,X_g(x)\rangle \big)\big(|c_\delta|^{1/2} Z+\langle \gamma e_{i'},X_g(x')\rangle\big)\big].
% \end{multline} 
%One can then apply Lemma \eqref{kah2} with $d=|\mathcal{N}|$, $\tilde{X}^i(x)\coloneqq |c_\delta|^{1/2} Z+\langle \gamma e_{i'},X_g(x)\rangle$, $X^i(x)\coloneqq |c_\delta|^{1/2} Z_i+Y_i(x)$ and $F(X^i,\dots,X^d)\coloneqq \prod_{i\in \mathcal{N}}  ( Z_{(x_j,\alpha_j)}^{\gamma e_{i_0}}  (B(x_j,\delta/2)))^{-   s_{i_0}}   \prod_{i\in \mathcal{N},i\not=i_0}         ( Z_{(x_j,\alpha_j)}^{\gamma e_i}  (B_i))^{-   s_i}  $
\qed

\section{\green{Some detailed perspectives}}
Toda CFTs provide natural extensions of Liouville CFT with a higher level of symmetry in addition to the Weyl anomaly which encodes the local conformal structure. In this document we have constructed Toda CFTs but we have not really shed light on where this W-symmetry does appear in the model and how useful it can be. We \green{detail} below some interesting questions related to this observation.
\subsection{W-symmetry and local conformal structure of Toda theories}\label{sec:sym}
\subsubsection{Local conformal structure of Toda Field Theories}
\iffalse
Let us consider the first non-trivial extension of the Liouville theory, \textit{i.e.} when one works with two negatively correlated GFFs---like in the construction of the $\mathfrak{sl}_3$ Toda theory. Then we can create a one-parameter family of QFTs by changing the covariance between these two GFFs, Lie algebras arising only for some special values of this parameter, that is when the covariance matrix coincides with the Cartan matrix of some Lie algebra. More explicitly one can assume to be working with a pair of GFFs whose covariance matrix is given by
\[
	\E[X^g_i(x) X^g_j(y)]=A_{i,j}   G_g(x,y),
\]
but where $A=\begin{pmatrix}
2 & c\\
c & 2\\
\end{pmatrix}$ with $c\in(-2,2)$ would no longer be the Cartan matrix of some semisimple Lie algebra (except for the values $c=0,-1$ corresponding to $\mathfrak{sl}_2\oplus\mathfrak{sl}_2$ and $\mathfrak{sl}_3$ Lie algebras). Interestingly these theories also enjoy conformal invariance thus define CFTs, but are not supposed to enjoy higher-spin symmetry. In this context it seems natural to wonder what is so specific about the theories defined via the $\mathfrak{sl}_3$ Lie algebra, and how can one see where W-symmetry does appear ?
These questions are being investigated in a work in progress.
More precisely it is expected that the existence of higher-spin currents that are \textit{holomorphic} is granted if and only if the covariance matrix of the GFFs is given by the Green kernel times the Cartan matrix of the $\mathfrak{sl}_3$ Lie algebra.
\fi

In Liouville CFT, the Weyl anomaly (combined with diffeomorphism invariance of the theory) is in some sense equivalent to the existence of a holomorphic current: the stress-energy tensor $T(z)$. The expression of this tensor can be obtained by 
formally \green{differentiating} the correlation function with respect to the metric
    \[
    \langle T_{\mu,\nu}(z)V_{\alpha_1}(z_1) \cdots V_{\alpha_k}(z_N)   \rangle_{g}\coloneqq 4\pi\frac{\partial}{\partial g^{\mu,\nu}}\langle V_{\alpha_1}(z_1) \cdots V_{\alpha_k}(z_N)   \rangle_{g},
    \]
and then setting $T\coloneqq T_{z,z}+\frac{c}{12}t$ where $t$ is explicit and depends on the background metric $g$. The first Ward identity 
\begin{equation}
\ps{T(z_0)\prod_{k=1}^NV_{\alpha_k}(z_k)}= \left(\sum_{k=1}^N \frac{\Delta_{\alpha_k}}{(z_0-z_k)^2}+\sum_{k=1}^N\frac{\partial_{z_k}}{z_0-z_k}\right)\ps{\prod_{k=1}^NV_{\alpha_k}(z_k)}
\end{equation}
and the asymptotic behaviour of the stress-energy tensor $T(z)\sim\frac1{z^4}$ near infinity (which comes from the fact that we have conformally mapped the sphere to the plane) usually ensure conformal covariance of the model. In a similar way Toda CFTs feature higher-spin currents $\bm {\mathrm W^{(k)}}(z)$ which should encode higher-spin symmetry via equations that take the same form as the Ward identity. As an application of our formalism it is possible to check that these identities hold (at least in the simplest $\mathfrak{sl}_3$ case) and study the properties of integrability provided by the W-symmetry. This has been recently carried out by the first author and Y. Huang in~\cite{Toda_OPEWV} where Ward identities for the $\mathfrak{sl}_3$ Toda CFT have been rigorously derived.

\subsubsection{Integrability for the $\mathfrak{sl}_3$ Toda CFT}
The structure of $W$-algebras is much more complicated that the Virasoro one (for instance higher-spin tensors feature higher derivatives and the commutation relations of their modes are no longer linear) and the tools used to prove integrability of Liouville theory come with additional complications. However and based on the exploitation of the symmetries of the model a first step towards integrability of the $\mathfrak{sl}_3$ Toda CFT has been carried out by the first author and Huang in~\cite{Toda_OPEWV} where a differential equation for some four-point correlation functions has been derived. This differential equation is the starting point for a mathematically rigorous computation of certain three-point correlation functions in the $\mathfrak{sl}_3$ Toda CFT, predicted in the physics literature~\cite[Equation (14)]{FaLi0}. This question is being investigated by the first author.

\subsection{The semi-classical limit and Toda equations in W-geometry}\label{semi-classical}
Let us comment here on the geometrical signification of \green{$\mathfrak{sl}_n$ Toda CFTs}. Their path integral formulation rely on the action functional \eqref{action} that corresponds to the quantization (\textit{i.e.} where we have introduced a coupling constant and considered an appropriate renormalization of the field and cosmological constants) of the action
\[
    S^c_T(\green\phi,g)\coloneqq  \frac{1}{4\pi} \int_{\Sigma}  \Big (  \langle\partial_g \green\phi(x), \partial_g \green\phi(x) \rangle_g   +2 R_g \langle \rho, \green\phi(x) \rangle +2\Lambda \sum_{i=1}^{n-1} e^{\langle e_i,\green\phi(x) \rangle}   \Big)\,{\rm v}_{g}(dx),
\]
whose critical point is given by the solution of the Toda equation \eqref{Toda_equ}. Such a critical point exists and is unique as soon as $\Sigma$ admits a metric for which $R_g$ is negative and constant; when the surface has the topology of the sphere or the torus one may need the field to have certain logarithmic singularities in order for such a problem to admit a unique solution. 

In the simplest case where $n-1=1$ the equation corresponds to the Liouville equation that describes (the conformal factor of) Riemannian metrics with constant negative curvature equal to $-\Lambda$ within the conformal class of $(\Sigma,g)$. In general, such an interpretation remains possible but instead of working in the setup of conformal geometry the good framework to consider is the one of \textit{W-geometry}. The interested reader may find more details on $W$-geometries in the work of Gervais and Matsuo \cite{GeMa92}. One of its important features is that, in a way similar to the fact that $W$-algebras admit the Virasoro algebra as a subalgebra, $W$-geometries in some sense contain two-dimensional Riemannian geometry. To be more specific,  the notion of \textit{$W$-surface} associated with the pair $(\S^2,\mathfrak{sl}_n)$ may be defined as a holomorphic embedding $(f(z), \bar f(\bar z))$ from $\S^2\simeq \C \mathrm P^1$ into the complex projective space $\C \mathrm P^{n-1}$ equipped with its Fubini-Study metric $g^{FS}$. This embedding naturally comes along
\footnote{For these vectors to be everywhere well-defined one may assume that some Wronskian that can be associated to the embedding does not vanish.}
with a family of $2(n-1)$ orthogonal vectors $(\bm{v_1},\cdots,\bm{v_{n-1}};\bm{\bar v_1},\cdots,\bm{\bar v_{n-1}})$ on $\C \mathrm P^{n-1}$ such that $\bm{v_1},\bm{\bar v_1}$ are tangent while the others are normal---to the embedded surface. Renormalizing these vectors one may then define a \textit{Frenet-Serret frame} $(\bm{e_1},\cdots,\bm{e_{n-1}};\bm{\bar e_1},\cdots,\bm{\bar e_{n-1}})$, that is an orthonormal basis of the projective space that is \lq\lq adapted" to the form of this embedding. Now on the one hand, the (covariant) derivatives of these vectors can be related to the Riemann curvature tensor $\mathcal{R}$ of $\C\mathrm{P}^{n-1}$ via the Gauss-Codazzi equations as follows:
\begin{equation}\label{Gauss-Codazzi}
    [\nabla,\bar\partial]\bm{e_k}=\sum_{l=1}^{n-1}\mathcal{R}(\partial f,\bar\partial h)_k^l\bm{e_l}.
\end{equation}
On the other hand and using the explicit expression of the vectors $(\bm{v_l},\bm{\bar v_l})_{1\leq l\leq n-1}$ we can get an explicit expression for the $[\nabla,\bar\partial]\bm{e_k}$ in terms of the $\ps{\varphi,e_i}\coloneqq -\ln\tau_i$, $1\leq i\leq n-1$, where $\norm{v_i}^2=\tau_i\tau_{i+1}$ and $\tau_1,\tau_{n}$ are determined by the embedding.
For these explicit computations to be consistent with the Gauss-Codazzi equations \eqref{Gauss-Codazzi} we need $\varphi$ to be a solution to the $\mathfrak{a}_n$-valued Toda equation:
\begin{align}\label{Toda_equ_2}
    \Delta\varphi+\sum_{i=1}^{n-1}e_ie^{\ps{e_i,\varphi}}=0 \quad\text{on }\C
\end{align}
with the precribed asymptotic
\begin{align}
    \varphi\sim-4\rho\ln\norm{z}\quad\text{as }z\to\infty.
\end{align}
Therefore Toda equations can in some sense be interpreted as compatibility equations for a holomorphic embedding of a two-sphere into a complex projective plane; yet there is another interpretation of these equations as \textit{zero-curvature conditions} for the second and third fundamental forms of the embedding, a fact which was already noticed in~\cite{LS79}. In brief we may establish a correspondence between solutions of the Toda equation \eqref{Toda_equ_2} and certain holomorphic embeddings from $\C\mathrm{P}^1$ to $\C\mathrm{P}^n$, which in some sense generalizes the problem of \textit{Uniformisation of Riemann surfaces}. Note that here and unlike what has been studied in this document the Toda equation is related to the problem of finding a metric with constant \textit{positive} curvature. The study is slightly more involved in the negative curvature case (since it requires the addition of singular points in the embedding) and will not be detailed in the present document. Details can be found in the aforementioned article by Gervais and Matsuo~\cite{GeMa92} and subsequent works.

From the quantum theory viewpoint, since we have constructed Toda CFTs via a quantization of this classical model, it is natural to expect that when the coupling constant $\gamma$ (which characterizes the level of randomness) is taken to zero ---and under the appropriate renormalizations $\phi^*=\gamma\phi$, $\alpha_k=\frac{\chi_k}{\gamma}$ and $\mu_i=\frac{\Lambda}{\gamma^2}$ for fixed $\chi_k$ and $\Lambda>0$--- we recover the solution of the classical Toda equation \eqref{Toda_equ} (when it exists and is unique). \green{Performing such a semi-classical analysis has been succesfully done by H. Lacoin and the last two authors in \cite{LRV19} for Liouville CFT, where a large deviation principle has been discovered}; it should be reasonable to expect that the result extends for the general $\mathfrak{sl}_n$ Toda CFTs on the sphere.

%%%%%%%%%%%%%%%%%%%%%%%%%%%%%%%%%%%%%%%%%%%%%%%%%%%%%%%%%%%%%%%%%%%%%%%%%%%%%%%%%%%%%%%%%%%%%%%%%%%%%%%%%%%%%%%%%%%%%%%%%%%%%%%%%%%%%%%%%%%%%%%%     

\section{Appendix}
 
 In the appendix, we gather rather general and classical results on Gaussian vectors: first comparison lemmas and then a statement of the Girsanov theorem.
% \begin{lemma}
% Let $F$ be some smooth function defined on $(\R^n)^d$ with at most polynomial growth at infinity for $F$ as well as for its derivatives up to order $2$. Assume that for $(x^1, \cdots, x^d) \in (\R^n)^d$ (where $x^i= (x^i_1 , \cdots, x^i_n)$) the folloxing inequality holds for all $i \not = j$ and for all $k,k'$
% \begin{equation*}
% \frac{\partial^2 F}{\partial x^i_k \partial x^j_{k'} } \geq 0
% \end{equation*}
% Then let $X=(X^1, \cdots, X^d)$ be a centered Gaussian vector  in $(\R^n)^d$ such that 
% for all $i \not = j$ and for all $k,k'$
% \begin{equation*}
% E[  X^i_k X^j_{k'}   ] \leq 0
% \end{equation*}
%Let  $\tilde{X}=(\tilde{X}^1, \cdots, \tilde{X}^d)$ be a centered Gaussian vector  in $(\R^n)^d$ such that $\tilde{X}^i $ has same distribution as $X^i$ for all $i$ and the vectors $\tilde{X}^i $ are independent. Then the following inequality holds 
%\begin{equation*}
%\E[    F(  X^1, \cdots, X^d  )  ] \leq \E[    F(  \tilde{X}^1, \cdots, \tilde{X}^d  )  ]
%\end{equation*}
% \end{lemma}
 \begin{lemma}\label{kah2}
 Let $F$ be some smooth function defined on $(\R^n)^{\green r}$ with at most polynomial growth at infinity for $F$ as well as for its derivatives up to order $2$. Assume that for $(x^1, \cdots, x^{\green r}) \in (\R^n)^{\green r}$ (where $x^i= (x^i_1 , \cdots, x^i_n)$) the following inequalities hold: 
 \begin{equation*}
\text{for all $i \not = j$ and $k,k'$}\quad \frac{\partial^2 F}{\partial x^i_k \partial x^j_{k'} } \geq 0.
 \end{equation*}
Let $X\coloneqq (X^1, \cdots, X^{\green r})$ and $\tilde X\coloneqq (\tilde{X}^1, \cdots, \tilde{X}^{\green r})$ be two centered Gaussian vectors  in $(\R^n)^{\green r}$ such that \\
1)  for all $i \not = j$ and $k,k'$
 \begin{equation*}
 \E[  X^i_k X^j_{k'}   ] \leq   \E[  \tilde{X}^i_k\tilde{X}^j_{k'}   ].
 \end{equation*}
2)  for all $i $,  $X^i$ as the same law as $\tilde{X}^i$.
 
Then the following inequality holds:
\begin{equation*}
\E[    F(  X^1, \cdots, X^{\green r}  )  ] \leq \E[    F(  \tilde{X}^1, \cdots, \tilde{X}^{\green r}  )  ].
\end{equation*}
 \end{lemma}

\proof
For $t\in [0,1]$, we set $X_t= \sqrt{t} X+\sqrt{1-t} \tilde{X}$, where $X$ and $\tilde{X}$ are independent, and 
\begin{equation*}
G(t)= \E[  F(   X_t ) ].
\end{equation*}
By using Gaussian integration by parts, we get   the following relation
\begin{align*}
G'(t) & = \frac{1}{2} \sum_{i=1}^d \sum_{k=1}^n  \E \left [  \frac{\partial F} {\partial x^i_k } (X_t) (  \frac{1}{\sqrt t} X^i_k - \frac{1}{\sqrt {1-t}} \tilde{X}^i_k  ) \right ]  \\
& = \frac{1}{2} \sum_{i=1}^d \sum_{k=1}^n   \sum_{i'=1}^d \sum_{k'=1}^n \E \left [ \frac{\partial^2 F} {\partial x^i_k \partial x^{i'}_{k'} } (X_t)  \right ] \E \left [( \sqrt t X^{i'}_{k'} +\sqrt {1-t} \tilde{X}^{i'}_{k'}  )    (  \frac{1}{\sqrt t} X^i_k - \frac{1}{\sqrt {1-t}} \tilde{X}^i_k  ) \right ]    \\
& = \frac{1}{2} \sum_{i=1}^d \sum_{k=1}^n   \sum_{i' \not =i}^d \sum_{k'=1}^n \E \left [ \frac{\partial^2 F} {\partial x^i_k \partial x^{i'}_{k'} } (X_t) \right   ] \big(\E [X^{i'}_{k'} X^i_k ] -\E [\tilde{X}^{i'}_{k'} \tilde{X}^i_k ] \big) \\
& \leq 0.
\end{align*}
Therefore $G(1)\leq G(0)$.
\qed    

 \begin{corollary}\label{comparison}
 Let \green{$H$} be some smooth function defined on $(\R_+)^{\green r}$ with at most polynomial growth at infinity for $\green H$ as well as for its derivatives up to order $2$ and consider a partition $P_1,\dots, P_m$ of the set $\{1,\dots,{\green r}\}$. Assume that  for $(x_1, \cdots, x_{\green r}) \in \R_+^{\green r}$,   the following inequality holds for all $\green{s,s'}\in \{1,\dots,m\}$ with $\green{s\not=s'}$, all $i\in P_{\green s}$ and all $j\in P_{\green s'}$   
 \begin{equation*}
 \frac{\partial^2 \green H}{\partial x_i  \partial x_j } \geq 0.
 \end{equation*}
Further assume that $X^1, \cdots, X^{\green r}$ is a family of continuous centered Gaussian fields respectively defined over  domains $D_i\subset \R^n$ (for $i=1,\dots,{\green r}$) such that for all $\green{ s,s'}\in \{1,\dots,m\}$ with $\green{s\not=s'}$, all $i\in P_{\green s}$ and all $j\in P_{\green s'}$   
 \begin{equation*}
\forall x\in D_i,\forall x'\in D_j,\quad \E[  X^i(x) X^j(x')   ] \leq 0.
 \end{equation*}
Let  $\tilde{X}=(\tilde{X}^1, \cdots, \tilde{X}^{\green r})$ be another family of continuous centered Gaussian fields such that:\\ 
1)  for all $\green s=1,\dots,m$, $ (\tilde{X}^i)_{i\in P_{\green s}} $ has same distribution as $(X^i)_{i\in P_{\green s}} $ .\\
2)  the families $(\tilde{X}^i)_{i\in P_1},\dots ,(\tilde{X}^i)_{i\in P_m}$  are independent. 

Eventually, let $f_1,\dots,f_{\green r}$ be a family of positive functions each of which respectively defined on $D_i$. For $i=1,\dots, {\green r}$, we set
$$M^i\coloneqq \int_{D_i} e^{X^i(x)-\frac{1}{2}\E[X^i(x)^2] }f_i(x)\,dx\quad \text{ and }\quad \tilde M^i\coloneqq \int_{D_i} e^{\tilde X^i(x)-\frac{1}{2}\E[X^i(x)^2] }f_i(x)\,dx.$$
Then the following inequality holds 
\begin{equation*}
\E[    \green H( M^1, \cdots, M^{\green r}  )  ] \leq \E[    H(  \tilde{M}^1, \cdots, \tilde{M}^{\green r}  )  ].
\end{equation*}
 \end{corollary}

\proof Up to a discretization of the fields, it suffices to apply Lemma \ref{kah2} with 
$$\green{F(X^1,\dots,X^{\green r})}\coloneqq \green H\left(\sum_{k_1}p^1_{k_1}e^{\gamma X^1_{k_1}-\frac{\gamma^2}{2}\E[(X^1_{k_1})^2]},\cdots,\sum_{k_{\green r}}p^{\green r}_{k_{\green r}}e^{\gamma X^{\green r}_{k_{\green r}}-\frac{\gamma^2}{2}\E[(X^{\green r}_{k_{\green r}})^2]}\right)$$
for some nonnegative numbers $p^i_{k_i}$ obtained by discretizing $f_i$ over $D_i$.
\qed

After these two comparison lemmas we recall the statement of the Girsanov \green{(or Cameron-Martin) theorem, see \emph{e.g.} Chapter VIII in~\cite{RY91}}. It  can be adapted in a way similar to the above proof via a regularization procedure in order to fit to the GFF we have considered throughout the present document:
\begin{theorem}[Girsanov theorem]\label{Girsanov}
Let $D$ be a subdomain of $\C$ and \[
(\bm X(x))_{x\in D}\coloneqq (X_1(x),\cdots,X_{n-1}(x))_{x\in D}\]
be a family of smooth centered Gaussian field; also consider $Z$ any Gaussian variable belonging to the $L^2$ closure of the subspace spanned by $(\bm X(x))_{x\in D}$. Then, for any bounded functional $F$ over the space of continuous functions one has that
\[
\expect{e^{Z-\frac{\expect{Z^2}}{2}}F(\bm X(x))_{x\in D}}=\expect{F\left(\bm X(x)+\expect{Z\bm X(x)}\right)_{x\in D}}.
\]
\end{theorem}

%%%%%%%%%%%%%%%%%%%%%%%%%%%%%%%%%%%%%%%%%%%%%%%%%%%%%%%
\bibliographystyle{plain}
\bibliography{biblio}

\begin{thebibliography}{10}

\bibitem{Arakawa_intro}
T.~Arakawa.
\newblock {\em Perspectives in Lie Theory}, chapter Introduction to W-Algebras
  and Their Representation Theory, pages 179--250.
\newblock Springer International Publishing, Cham, 2017.

\bibitem{BPZ}
A.A. Belavin, A.M. Polyakov, and A.B. Zamolodchikov.
\newblock Infinite conformal symmetry in two-dimensional quantum field theory.
\newblock {\em Nuclear Physics B}, 241(2):333 -- 380, 1984.

\bibitem{Borcherds}
R.~Borcherds.
\newblock {Vertex algebras, Kac-Moody algebras, and the Monster}.
\newblock In {\em {Proceedings of the National Academy of Sciences of the
  United States of America}}, volume~83, pages 3068--3071, 1986.

\bibitem{Toda_OPEWV}
B.~Cercl\'e and Y.~Huang.
\newblock {Ward identities in the $\mathfrak{sl}_3$ Toda field theory}.
\newblock {\em arXiv preprint}, arXiv:2105.01362, 2021.

\bibitem{DKRV}
F.~David, A.~Kupiainen, R.~Rhodes, and V.~Vargas.
\newblock Liouville {Q}uantum {G}ravity on the {R}iemann {S}phere.
\newblock {\em Communications in Mathematical Physics}, 342:869, 2016.

\bibitem{DRV16}
F.~David, R.~Rhodes, and V.~Vargas.
\newblock Liouville quantum gravity on complex tori.
\newblock {\em Journal of Mathematical Physics}, 57(2):022302, 2016.

\bibitem{DDDF}
J.~{Ding}, J.~{Dub{\'e}dat}, A.~{Dunlap}, and H.~{Falconet}.
\newblock {Tightness of Liouville first passage percolation for $\gamma \in
  (0,2)$}.
\newblock {\em {Publications mathématiques de l'IH\'ES}}, {132}:353--403,
  2020.

\bibitem{DO94}
H.~Dorn and H.-J. Otto.
\newblock {Two- and three-point functions in Liouville theory}.
\newblock {\em Nuclear Physics B}, 429(2):375 -- 388, 1994.

\bibitem{dubedat}
J.~Dub\'edat.
\newblock {SLE and the Free Field: partition functions and couplings}.
\newblock {\em {Journal of the AMS}}, 22 (4):995--1054, 2009.

\bibitem{DFGPS}
J.~Dub\'edat, H.~Falconet, E.~Gwynne, J.~Pfeffer, and X.~Sun.
\newblock {Weak LQG metrics and Liouville first passage percolation}.
\newblock {\em Probability Theory and Related Fields}, 178:369--436, 2020.

\bibitem{DMS14}
B.~{Duplantier} and S.~{Miller}, J.~{Sheffield}.
\newblock {Liouville quantum gravity as a mating of trees}.
\newblock {\em arXiv preprint}, {arXiv}:1409.7055, 2014.

\bibitem{FaLi0}
V.A. Fateev and A.V. Litvinov.
\newblock {On differential equation on four-point correlation function in the
  Conformal Toda Field Theory}.
\newblock {\em JETP Lett.}, 81:594--598, 2005.

\bibitem{FaLu}
V.A. Fateev and S.~L. Lukyanov.
\newblock {The Models of Two-Dimensional Conformal Quantum Field Theory with
  Z(n) Symmetry}.
\newblock {\em Int. J. Mod. Phys. A}, 3:507, 1988.

\bibitem{FaZa2}
V.A. Fateev and A.B. Zamolodchikov.
\newblock {Parafermionic Currents in the Two-Dimensional Conformal Quantum
  Field Theory and Selfdual Critical Points in Z(n) Invariant Statistical
  Systems}.
\newblock {\em Sov. Phys. JETP}, 62:215--225, 1985.

\bibitem{FaZa}
V.A. Fateev and A.B. Zamolodchikov.
\newblock {Conformal quantum field theory models in two dimensions having Z3
  symmetry}.
\newblock {\em Nuclear Physics B}, 280:644 -- 660, 1987.

\bibitem{FORTW}
L.~Fehér, L.~O'Raifeartaigh, P.~Ruelle, I.~Tsutsui, and A.~Wipf.
\newblock {On Hamiltonian reductions of the Wess-Zumino-Novikov-Witten
  theories}.
\newblock {\em Physics Reports}, 222(1):1 -- 64, 1992.

\bibitem{FLM89}
I.~Frenkel, J.~Lepowsky, and A.~Meurman.
\newblock {\em {Vertex Operator Algebras and the Monster}}.
\newblock Academic Press, 1989.

\bibitem{FdV}
H.~Freudenthal and H.~de~Vries.
\newblock {\em Linear lie groups}.
\newblock Pure and Applied Mathematics. Academic Press, 1969.

\bibitem{GeMa92}
J.-L. Gervais and Y.~Matsuo.
\newblock {W geometries}.
\newblock {\em Phys. Lett. B}, 274:309--316, 1992.

\bibitem{GKRV}
C.~Guillarmou, A.~Kupiainen, R.~Rhodes, and V.~Vargas.
\newblock {Conformal bootstrap in Liouville Theory}.
\newblock {\em arXiv preprint}, arXiv:2005.11530, 2020.

\bibitem{GRV16}
C.~Guillarmou, R.~Rhodes, and V.~Vargas.
\newblock {Polyakov's formulation of $2d$ bosonic string theory}.
\newblock {\em Publications Math\'ematiques de l'IH\'ES}, 130:111--185, 2016.

\bibitem{GM20}
E.~Gwynne and J.~Miller.
\newblock {Existence and uniqueness of the Liouville quantum gravity metric for
  \\$\gamma\in(0,2)$}.
\newblock {\em Inventiones mathematicae}, 223:213--333, 2021.

\bibitem{HRV16}
Y.~Huang, R.~Rhodes, and V.~Vargas.
\newblock Liouville quantum gravity on the unit disk.
\newblock {\em Ann. Inst. H. Poincar\'e Probab. Statist.}, 54(3):1694--1730, 08
  2018.

\bibitem{Hum72}
J.~Humphreys.
\newblock {\em {Introduction to Lie Algebras and Representation Theory}}.
\newblock Springer, Berlin, 1972.

\bibitem{JMO}
M.~Jimbo, T.~Miwa, and M.~Okado.
\newblock {Solvable lattice models related to the vector representation of
  classical simple Lie algebras}.
\newblock {\em Communications in Mathematical Physics}, 116(3):507--525, 1988.

\bibitem{Kah}
J.-P. Kahane.
\newblock Sur le chaos multiplicatif.
\newblock {\em Annales des sciences math{\'{e}}matiques du Qu{\'{e}}bec}, 1985.

\bibitem{KRV_loc}
A.~Kupiainen, R.~Rhodes, and V.~Vargas.
\newblock Local {C}onformal {S}tructure of {L}iouville {Q}uantum {G}ravity.
\newblock {\em Communications in Mathematical Physics}, 2018.

\bibitem{KRV_DOZZ}
A.~Kupiainen, R.~Rhodes, and V.~Vargas.
\newblock {Integrability of Liouville theory: proof of the DOZZ formula}.
\newblock {\em Annals of Mathematics}, 191(1):81--166, 2020.

\bibitem{LRV19}
H.~Lacoin, R.~Rhodes, and V.~Vargas.
\newblock {The semiclassical limit of Liouville conformal field theory}.
\newblock {\em arXiv preprint}, arXiv:1903.08883, 2019.

\bibitem{LeG13}
J.-F. Le~Gall.
\newblock Uniqueness and universality of the {B}rownian map.
\newblock {\em Ann. Probab.}, 41(4):2880--2960, 07 2013.

\bibitem{LS79}
A.N. Leznov and M.V. Saveliev.
\newblock {Representation of zero curvature for the system of nonlinear partial
  differential equations $x_{\alpha,z\bar z}=\exp(kx)_\alpha$ and its
  integrability}.
\newblock {\em Letters in Mathematical Physics}, 3:489--494, 1979.

\bibitem{Mie13}
G.~Miermont.
\newblock The {B}rownian map is the scaling limit of uniform random plane
  quadrangulations.
\newblock {\em Acta Math.}, 210(2):319--401, 2013.

\bibitem{Sei90}
Seiberg N.
\newblock {Notes on Quantum Liouville Theory and Quantum Gravity}.
\newblock {\em Progress of Theoretical Physics Supplement}, 102:319--349, 03
  1990.

\bibitem{OPS88}
B.~Osgood, R.~R~Phillips, and P.~Sarnak.
\newblock {Extremals of determinants of Laplacians}.
\newblock {\em Journal of Functional Analysis}, 80(1):148 -- 211, 1988.

\bibitem{Pol81}
A.~Polyakov.
\newblock {Quantum Geometry of bosonic strings}.
\newblock {\em Physics Letters B}, 103:207:210, 1981.

\bibitem{RY91}
D.~Revuz and M.~Yor.
\newblock {\em {Continuous Martingales and Brownian Motion}}.
\newblock Springer, Berlin, 1991.

\bibitem{review}
R.~Rhodes and V.~Vargas.
\newblock Gaussian multiplicative chaos and applications: {A} review.
\newblock {\em Probab. Surveys}, 11:315--392, 2014.

\bibitem{schramm}
O.~Schramm.
\newblock Scaling limits of loop-erased random walks and uniform spanning trees
  [mr1776084].
\newblock In {\em Selected works of {O}ded {S}chramm. {V}olume 1, 2}, Sel.
  Works Probab. Stat., pages 791--858. Springer, New York, 2011.

\bibitem{She07}
S.~Sheffield.
\newblock Gaussian free field for mathematicians.
\newblock {\em {Probability Theory and Related Fields}}, 139:521, 2007.

\bibitem{ZZ96}
A.~Zamolodchikov and Al. Zamolodchikov.
\newblock {Conformal bootstrap in Liouville field theory}.
\newblock {\em {Nuclear Physics B}}, 477(2):577--605, 1996.

\bibitem{Za85}
A.~B. {Zamolodchikov}.
\newblock {Infinite additional symmetries in two-dimensional conformal quantum
  field theory}.
\newblock {\em Theoretical and Mathematical Physics}, 65(3):1205--1213,
  December 1985.

\end{thebibliography}
\end{document}